\documentclass[twocolumn]{aastex63} 

\usepackage[colorinlistoftodos]{todonotes}
\usepackage[flushleft]{threeparttable}
\usepackage{graphicx}

\let\Oldtodo\todo
\renewcommand{\todo}[1]{\Oldtodo[inline]{#1}}

\hypersetup{linkcolor=cyan,citecolor=cyan,filecolor=cyan,urlcolor=cyan}

\defcitealias{halmil26}{Paper I}
\defcitealias{halmil26b}{Paper II}

\usepackage{amsmath}
\usepackage{float}

\shorttitle{Tilted Stars in the Desert}
\shortauthors{Hallatt, Owen, \& Millholland}

\graphicspath{{./}{figures/}}

\begin{document}

\title{Revealing the Origin of Desert Dwellers via Stellar Obliquities}

\correspondingauthor{Tim Hallatt}
\email{thallatt@mit.edu}

\author[0000-0003-4992-8427]{Tim Hallatt}
\affil{MIT Kavli Institute for Astrophysics and Space Research, Massachusetts Institute of Technology, Cambridge, MA 02139, USA}

\author[0000-0002-4856-7837]{James E. Owen}
\affil{Imperial Astrophysics, Department of Physics, Imperial College London, Prince Consort Road, London SW7 2AZ, UK}

\author[0000-0003-3130-2282
]{Sarah Millholland}
\affil{Department of Physics, Massachusetts Institute of Technology, Cambridge, MA 02139, USA}
\affil{MIT Kavli Institute for Astrophysics and Space Research, Massachusetts Institute of Technology, Cambridge, MA 02139, USA}

\begin{abstract}

Observations suggest that the hot Neptune desert 
contains the remnants of destroyed gas giants. Recent theoretical work has shown that gas giant destruction via Roche lobe overflow (RLO) can indeed populate the desert with remnant planets, but only if mass transfer removes most of the planet's orbital angular momentum (``lossy" RLO). Motivated by the fact that stellar accretion naturally gives rise to such lossy RLO, in this Letter we examine how planet-to-star mass and angular momentum transfer manifests in the distribution of stellar obliquities.
We find that RLO tilts host stars into spin/orbit alignment (within a few ${\sim}$tens of degrees) regardless of initial conditions. Obliquity damping by RLO can only be reversed by the presence of misaligned companion planets within ${\lesssim}$2 au. While tides and mass transfer usually produce stellar spin up, host stars can also emerge from RLO slowly rotating if systems begin strongly retrograde; retrograde RLO reconciles theory with the anomalously slow rotation of the desert dweller host, LTT 9779. Predicted spin/orbit alignment may differentiate RLO from alternative giant planet destruction mechanisms, in particular hot Jupiter disruption during high eccentricity migration (which tends to produce broadly distributed stellar obliquities). We summarize other population-level predictions that can further distinguish RLO from high eccentricity migration. Our work suggests that follow-up obliquity measurements may reveal the formation pathways of desert dwellers, and potentially open a window into gas giants' exposed interiors.

\end{abstract}

\section{Introduction} \label{sec:intro}

The $\textit{Kepler}$ spacecraft revealed that  
hot Jupiters (${\gtrsim}8 \ R_{\oplus}, \ {\gtrsim}100 \ M_{\oplus}$) and super-Earths/sub-Neptunes (${\lesssim}4 \ R_{\oplus}, \ {\lesssim}10 \ M_{\oplus}$) inside orbital periods ${\lesssim}$3 days are wedged apart by a dearth of mid-sized planets \citep[e.g.][]{szakis11,beanes13}. The census of planets surrounding this ``sub-Jovian desert" has recently been brought under new light by $\textit{TESS}$ \citep[][see e.g. \citealt{jendiamat20,armlopadi20}]{ricwinvan15}. In contrast to $\textit{Kepler}$-era estimates that the desert was virtually empty \citep[hence the name ``desert"; e.g.][]{szakis11,mazholfai16}, $\textit{TESS}$ has revealed a number of planets residing squarely in this region previously thought uninhabitable \citep[``desert dwellers";][]{visbeh25,doyarmacu25}; though sparsely populated, ultrahot mid-sized planets $\textit{do}$ exist. The goal of this paper is to motivate and guide observations that may elucidate the origin of these unexpected desert dwellers.

The sub-Jovian desert records the migratory and mass loss histories of the hottest exoplanets. Opening inward from ${\sim}$3 days toward smaller masses/radii, the desert's bottom boundary may reflect atmospheric erosion of low-mass planets catalyzed by stellar irradiation \citep[e.g.][]{lunkjealb16,owelai18,hallee22}. Stretching upward in planet mass/radius toward shorter periods, the desert's top boundary likely traces the tidal disruption limit of giant planets during high eccentricity migration \citep[][a similar picture may also apply to the lower boundary following low-mass planets' differing mass/radius relation; \citealt{matkon16}]{owelai18}, with minimal contribution from atmospheric mass loss (\cite{ionpav18,visknugre22}, though see \cite{kurnak14} for a differing perspective). In-situ formation of hot Jupiters could also help carve the desert's upper boundary, with tidal orbital decay setting the innermost edge of the hot Jupiter population \citep[][]{baibat18}.

Individual $\textit{TESS}$ planet discoveries as well as completeness-corrected $\textit{TESS}$ occurrence rates \citep{cuiarmhad26} confirm a collection of planets residing within these boundaries. Motivated by the fact that desert dwellers orbit metal-rich stars indistinguishable from hot Jupiters \citep{visbeh25}, and boast high densities consistent with massive, stripped planetary cores \citep{armlopadi20,doyarmacu25}, \cite{halmil26} showed that the sub-Jovian desert can be backfilled by the remnants of tidally destroyed gas giants; in this picture, hot Jupiters deposit their stripped remains across the desert via Roche lobe overflow (RLO) driven by tidal orbital inspiral. If confirmed by observations, the \cite{halmil26} theory suggests that desert dwellers provide a window into gas giant interiors (which are otherwise impossible to study directly), and reveal the fates of (at least some) hot Jupiters. 

In order for hot Jupiter RLO to successfully reproduce desert dwellers, \cite{halmil26} found that mass transfer must be ``lossy": most of the orbital angular momentum carried by material leaving the planet cannot be returned to the orbit. This ``lossy" RLO contrasts with classical mass transfer theory, in which most of the donor's orbital angular momentum is returned via torque from an accretion disk \citep[e.g.][]{ver82,verrap88}. As we will show, lossy RLO naturally arises due to stellar magnetospheric accretion of angular momentum, preventing any back-reaction torque on the planet. In this Letter we clarify how lossy RLO driven by stellar accretion of angular momentum and mass can be tested observationally and differentiated from other theories. We argue that lossy RLO makes a clean prediction for the distribution of stellar obliquities \citep[the angle between stellar spin and orbital angular momenta;][]{albdawwin22} in the sub-Jovian desert. Specifically, we find that angular momentum/mass transfer robustly tilts host stars into alignment with planetary orbits (within a few ${\sim}$tens of degrees). 

Predicted spin/orbit alignment can differentiate RLO from other mechanisms populating the desert. The chief alternative mechanism that may conceivably populate the desert is tidal disruption during high eccentricity migration (HEM), whereby dynamical tides excited in the planet at periapse drive catastrophic mass loss \citep[][though see \citealt{welhannao25} for further discussion of how violent this process may be]{guiramlin11,liuguilin13}, leaving behind a remnant core due to its higher density. If hot Jupiters are destroyed during HEM, we may expect their vestiges to inherit broadly-distributed obliquities produced during HEM \citep[e.g. via von Zeipel/Lidov/Kozai oscillations;][]{fabtre07,stoandlai14}. The obliquity predictions reported in this work therefore serve as strong motivation for follow-up observations of desert dwellers to test the \cite{halmil26} theory. 

Our goal of providing clean obliquity predictions to test the origin of desert dwellers is akin to ongoing efforts focused on planets beyond the desert \citep[in the so-called ``ridge"; e.g.][]{bouattmall23,attia2023, Louden2024,  yeeyamste25}. To the best of our knowledge, there are currently no published obliquity (projected or true) measurements for planets in the sub-Jovian desert sample of \cite{halmil26}. It is our hope that this paper provides motivation to extend the census of close-in planet obliquities into the desert.

This Letter is organized as follows. Section \ref{subsec:OoM} establishes that stellar spin and hot Jupiter orbital angular momenta are comparable, indicating that angular momentum/mass transfer during RLO can strongly influence the rate and direction of stellar spin (and therefore obliquity). We detail our numerical setup in Section \ref{sec:methods}, before outlining how stellar obliquities are tilted into alignment in Section \ref{sec:obl}. Section \ref{sec:companion} clarifies how deviations from predicted obliquity damping can arise after RLO due to external companion planets, estimating their parameters to help guide observations. Sections \ref{sec:HEM} and \ref{sec:discussion} summarize and tabulate the ways that observations can confirm the \cite{halmil26} picture. We close in Section \ref{sec:conclusion} with a brief recap of our results. 

\begin{figure*}[t]
\centering
\includegraphics[width=0.8\textwidth]{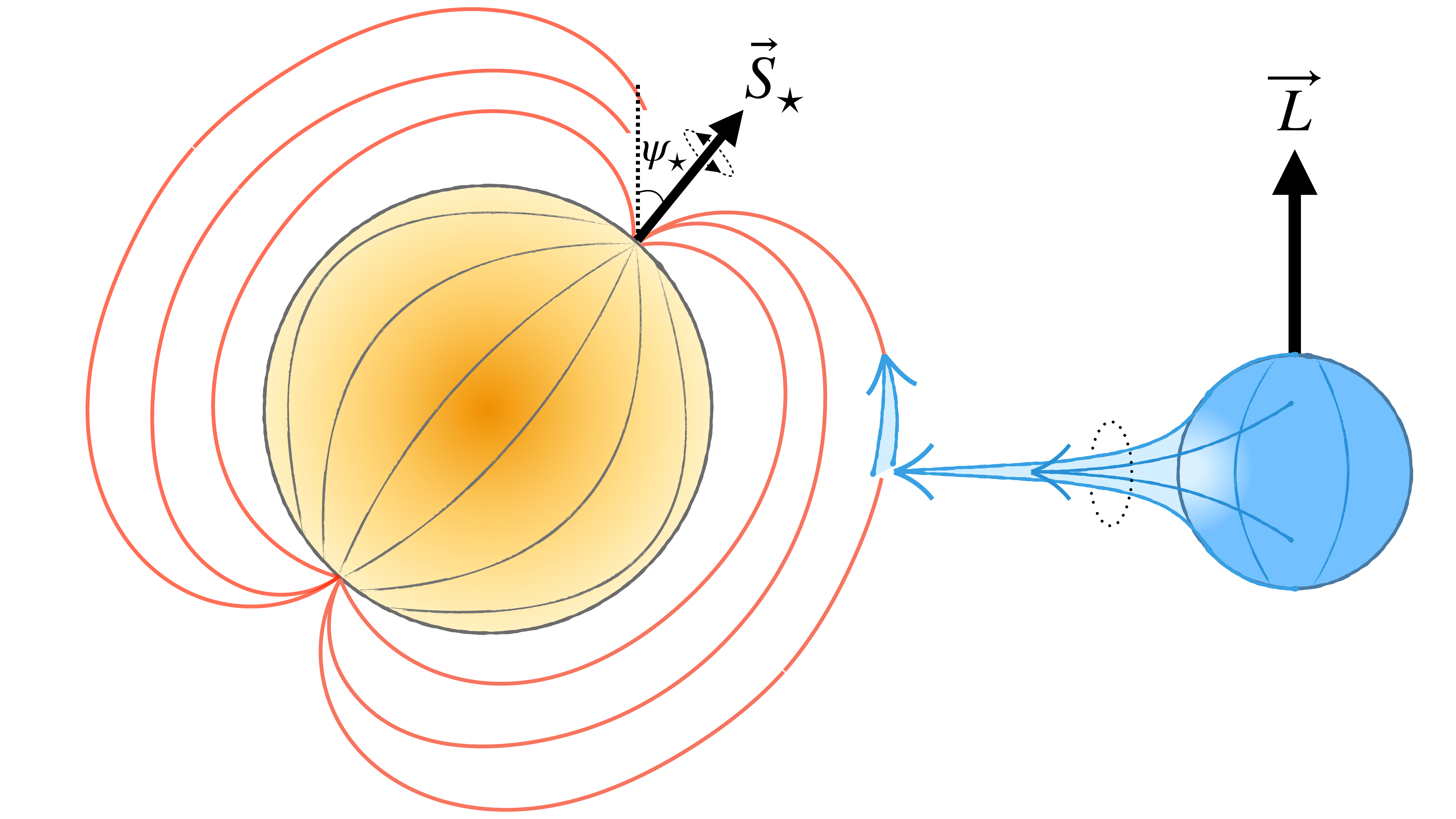}
\caption{Schematic of a hot Jupiter undergoing ``lossy" Roche lobe overflow. The host star spin angular momentum vector $\bf{S}_{\star}$ is misaligned with the planet's orbital angular momentum $\bf{L}$ by obliquity $\psi_{\star}$. Outflowing gas passes through a nozzle at the first Lagrange point (dotted ellipse), before coupling to the stellar magnetic field (in red) at the Alfvén radius; accreting gas is funneled along field lines onto the magnetic poles (assumed coincident with the rotation axis), preventing the formation of a disk. Angular momentum transfer during Roche lobe overflow shepherds $\bf{S}_{\star}$ into alignment with $\bf{L}$; obliquity is damped.
\label{figure:schematic}}
\end{figure*}

\section{Back of Envelope Expectation}\label{subsec:OoM}

Gas advected from a hot Jupiter to its host star during RLO transfers angular momentum in addition to mass. Upon impact at the stellar surface (or coupling to the stellar magnetic field), the orbital angular momentum of the accreting material is converted to stellar spin angular momentum. The degree to which this affects the stellar spin direction depends on the magnitude of angular momentum transfer. To guide our analysis, we begin this section by estimating typical orbital and stellar spin angular momenta of hot Jupiter systems. We will show that orbit/spin angular momenta are comparable, indicating that catastrophic RLO as explored by \cite{halmil26} 
can have a profound impact on the rate and direction of stellar spin. 

The ratio of orbital angular momentum, $L$, to stellar spin angular momentum, $S_{\star}$, reads (assuming a circular orbit and mass ratio $M_{\rm p}/M_{\star}{\ll}1$),

\begin{align}\label{equation:LS}
\begin{split}
    \frac{L}{S_{\star}}&=\frac{M_{\rm p}\sqrt{GM_{\star}a}}{\alpha_{\star}M_{\star}R^{2}_{\star}\Omega_{\star}}\\
    &\hspace{-0.6cm} \sim 6 \bigg(\frac{M_{\rm p}}{M_{\rm J}}\bigg)\bigg(\frac{M_{\star}}{M_{\odot}}\bigg)^{-1/3}\bigg(\frac{R_{\star}}{R_{\odot}}\bigg)^{-2}\bigg(\frac{P_{\rm orb}}{1 \ \rm d}\bigg)^{1/3}\bigg(\frac{P_{\star}}{20 \ \rm d}\bigg),
\end{split}
\end{align}

\noindent 
with $M_{\rm p}$ and $M_{\star}$ the planetary and stellar masses respectively, $a$ the orbital semi-major axis (converted to orbital period $P_{\mathrm{orb}}$ for convenience in Equation \ref{equation:LS}), $\alpha_{\star}{=}0.06$ the square of the stellar radius of gyration \citep[the square of ``Beta" under Table II.1 from][]{cla89}, and $R_{\star}$ and $\Omega_{\star}$ the stellar radius and spin frequency (converted to stellar spin period $P_{\star}$ in Equation \ref{equation:LS}) respectively. Our choices of $\alpha_{\star}$ and $P_{\star}$ are typical values for main sequence FGK stars \cite[][]{barweifri16}.

Counter to classical mass transfer theory, lossy RLO entails almost complete loss of angular momentum from the planet orbit. This loss likely arises due to stellar accretion of mass/momentum (see Section \ref{subsubsec:mt}). 
Equation \ref{equation:LS} informs us that, by transferring almost all of the planet's orbital angular momentum to the star, lossy RLO can change the stellar spin angular momentum significantly (unlike standard mass transfer). The task of our paper is to quantitatively assess the evolution in angular momenta using the detailed mass transfer calculations presented in \cite{halmil26}, tracking several additional processes that act in tandem with mass transfer. We next outline how we track the sum total changes in angular momenta.

\section{Methods}\label{sec:methods}

\subsection{Equations of Motion}

\subsubsection{Mass Transfer}\label{subsubsec:mt}

We define the orbital and stellar spin angular momentum vectors $\mathbf{L}{=}L\mathbf{\hat{L}}$ and $\mathbf{S_{\star}}=S_{\star}\mathbf{\hat{S}_{\star}}$, respectively (with $L$ and $S_{\star}$ defined in equation \ref{equation:LS}). Mass transfer in isolation changes the respective angular momenta via,

\begin{align}\label{equation:mt}
\begin{split}
    \frac{d\bf{L}}{dt}\bigg|_{\rm MT}&=\Gamma \frac{\dot{m}_{\rm RLO}}{M_{\rm p}}L \mathbf{\hat{L}},\\
    \frac{d\bf{S_{\star}}}{dt}\bigg|_{\rm MT}&=-\beta \frac{d\bf{L}}{dt}\bigg|_{\rm MT}\\
\end{split}
\end{align}

\noindent where $\dot{m}_{\rm RLO}{<}0$ is the planet's mass loss rate due to RLO \citep[computed as in][]{halmil26} and $L/M_{\rm p}$ is the specific orbital angular momentum. Following \cite{halmil26} we parameterize how much orbital angular momentum is returned to the planet orbit \citep[e.g. via tidal torque from accreting gas;][]{priwad85} by the constant $\Gamma$; in this setup, $\Gamma{=}0$ yields conservation of orbital angular momentum whereas $\Gamma{=}1$ reflects complete loss. As in \cite{halmil26}, we take $\Gamma{=}0.85$ as our fiducial value (constituting ``lossy" RLO; mass transfer results are unchanged for $0.65{\lesssim}\Gamma{\lesssim}0.95$). Lastly, the constant $\beta$ reflects the fraction of angular momentum lost from the orbit that successfully  accretes onto the host star; perfect transfer yields $\beta{=}1$, while total loss of angular momentum from the system produces $\beta{=}0$. 

\begin{deluxetable*}{CCCCCCc}\label{tab:fiducial_table}
\tablecaption{Initial conditions used in our Monte Carlo stellar obliquity exploration.} 
\label{tab:parameters}
\tablecolumns{5}
\tablewidth{0pt}
\tablehead{
\colhead{Parameter} &
\colhead{Definition} &
\colhead{Value} & 
\colhead{Units} & 
}
\startdata
$M_{\rm p}$ & \rm initial \ planet \ mass & \{1.5{\times}10^{2}, \ 3{\times}10^{2}, \ 9{\times}10^{2}\} & M_{\oplus} \\
$M_{\rm core}$ & \rm \ planet \ core \ mass & 20 & M_{\oplus} \\
$a$ & \rm initial \ semi${-}$major \ axis & 0.03 & \rm au \\
$Q_{\star}$ & \rm \ stellar \ tidal \ quality \ factor & \{10^4, \ 5{\times}10^{4}, \ 10^5\} & \nodata \\
$P_{\star}$ & \rm initial \ stellar \ rotation \ period & \{5, \ 30\} & \rm{days} \\
$\Gamma$ & \rm orbital \ angular \ momentum \ loss \ fraction & 0.85 & \nodata \\
$\beta$ & \rm stellar \ angular \ momentum \ accretion \ fraction & \{0, \ \beta(R_{\rm A}), \ 1\} & \nodata \\
$\cos \psi_{\star}$ & \rm cosine \ of \ stellar \ obliquity & \rm uniform \ in \ [-1,1]  & \nodata \\
$t$ & \rm system \ age & 10^{9} & \rm yr \\
\\
\enddata
\end{deluxetable*}

RLO naturally occurs under high degrees of angular momentum loss since 
the stellar magnetic field prevents gas from forming an accretion disk. Analogous to strongly magnetized ``polar"/AM Herculis cataclysmic variables \citep[e.g.][]{cam84,kinfrawhi90,cro90}, outflowing gas coupled to magnetic field lines will accrete directly onto the stellar surface instead of spreading into a disk (Figure \ref{figure:schematic}). Gas couples to the stellar magnetic field at the Alfvén radius, approximately located at \citep[e.g.][]{prires72,lampetpin73}:

\begin{equation}\label{equation:ralf}
    R_{\rm A}=\bigg(\frac{\mathcal{M}^{4}_{\rm B}}{8}\frac{1}{\dot{m}^{2}_{\rm RLO}GM_{\star}}\bigg)^{1/7},
\end{equation}

\noindent in which we have assumed a dipole magnetic field $B(r){=}\mathcal{M}_{\rm B}/r^{3}$ with magnetic moment $\mathcal{M}_{\rm B}{=}B_{\star}R^{3}_{\star}$ (where $B_{\star}{=}10 \ \rm G$ is the stellar magnetic field at $R_{\star}$; our results are not sensitive to $B_{\star}$). Without an accretion disk, the only angular momentum returned to the orbit arises due to the lower angular momentum of material leaving the planet from the first Lagrange point than at the planet location; for a mass ratio $M_{\rm p}/M_{\star}{\sim}10^{-3}$, $\Gamma{\sim}0.85$ \citep[][]{jiaspr17}.


Equation \ref{equation:ralf} provides an estimate of how much angular momentum is accreted by the star. Accreting material from $R_{\rm A}$ will yield a torque $d\mathbf{S_{\star}}/dt{=}{-}\dot{m}_{\rm RLO}\sqrt{G M_{\star}R_{\rm A}}\hat{\bf{L}}$, so that $\beta{=}\texttt{min}[\sqrt{R_{\rm A}/(\Gamma^{2}a)},1]$ (taking an upper limit $\beta{=}1$ to ensure momentum conservation, and a lower limit $\beta{=}\sqrt{R_{\odot}/(\Gamma^{2}a)}$). This parametrization is physically reasonable; $R_{\rm A}{=}a$ yields $\beta{=}1$, corresponding to ``direct impact accretion" in which material leaving L1 is immediately swept along field lines to the stellar surface. Meanwhile $R_{\rm A}{=}R_{\rm \odot}$ gives a lower limit of $\beta{\sim}0.5$ (for typical $a{\sim}0.02$ au). We verify in Section \ref{sec:obl} that varying $\beta{\in}[0.5,1]$ does not significantly change the conclusions of this paper.

\subsubsection{Photoevaporation}

Planetary mass loss via photoevaporation (``PE") driven by stellar soft X-ray (0.1${\lesssim}h\nu{\lesssim}$ 1-2 keV) and EUV ($h\nu{\geq}13.6$ eV) emission also carries away the planet's orbital angular momentum \citep[][]{valrapras15}. Assuming an isotropic wind, the angular momentum loss rate is:

\begin{equation}
    \frac{d\mathbf{L}}{dt}\bigg|_{\rm PE}=\frac{\dot{m}_{\rm PE}}{M_{\rm p}}L\mathbf{\hat{L}},
\end{equation}

\noindent with $\dot{m}_{\rm PE}{<}0$ the photoevaporative mass loss rate (computed as in \cite{halmil26}, using the energy limited approximation). While PE has negligible impact on hot Jupiters \citep[e.g.][]{murchimur09,owejac12}, it can affect remnant hot Neptunes' atmospheres \citep{halmil26}.

\subsubsection{Tidal Dissipation}\label{subsubsec:tides}

In addition to mass transfer, angular momentum is exchanged between orbit and stellar spin due to tides raised on the host star by the planet. Following \cite{halmil26} we assume the planet has zero planetary obliquity and eccentricity such that only stellar tides operate. We adopt for simplicity the equilibrium, weak friction model of \cite{hut81}; in vector notation, the tidal equations of motion read \citep[e.g.][]{matpearas10,lai12},

\begin{align}\label{equation:tides}
\begin{split}
    \frac{d\mathbf{S_{\star}}}{dt}\bigg|_{\rm tid}&=\mathcal{T}_{\parallel}\mathbf{\hat{S}_{\star}}+\mathcal{T}_{\perp}\mathbf{\hat{x}},\\
    \frac{d\mathbf{L}}{dt}\bigg|_{\rm tid}&=-\frac{d\mathbf{S_{\star}}}{dt}\bigg|_{\rm tid},\\
\end{split}
\end{align}

\noindent with $\mathcal{T}_{\parallel}$ the tidal component along the stellar spin direction and $\mathcal{T}_{\perp}$ the tidal component perpendicular to the stellar spin in the plane of $\bf{\hat{S}_{\star}}$ and $\bf{\hat{L}}$ (i.e. $\mathbf{\hat{x}}{=}{-}\mathbf{\hat{S}_{\star}}{\times}(\mathbf{\hat{S}_{\star}}{\times}\mathbf{\hat{L}})$).\footnote{Adopting a more sophisticated tidal model would not affect our conclusions.} The tidal torque components are,

\begin{align}
\begin{split}
    \mathcal{T}_{\parallel}&=\frac{3}{2}k_{2,\star}\delta t_{\star}\mathcal{T}_{0}\bigg[2\Omega\cos\psi_{\star}-(1+\cos^{2}\psi_{\star})\Omega_{\star}\bigg],\\
    \mathcal{T}_{\perp}&=\frac{3}{2}k_{2,\star}\delta t_{\star}\mathcal{T}_{0}\sin\psi_{\star}\bigg[2\Omega-\Omega_{\star}\cos\psi_{\star}\bigg],\\
\end{split}
\end{align}

\noindent with $k_{2,\star}{=}0.028$ the star's Love number \citep[twice the apsidal motion constant of 0.014 for a Sun-like star;][]{cla23}, $\delta t_{\star}{=}1/(2\Omega Q_{\star})$ the tidal lag time with $Q_{\star}$ the tidal quality factor (we explore $Q_{\star}{\in}\{10^{4},5{\times}10^{4},10^{5}\}$; variations in $Q_{\star}$ affect the inspiral time but do not strongly affect the mass transfer process) 
and $\Omega{=}\sqrt{GM_{\star}/a^{3}}$ the orbital frequency, $\mathcal{T}_{0}{=}GR^{5}_{\star}(M_{\rm p}/a^{3})^{2}$, and the obliquity $\psi_{\star}$ defined via $\cos \psi_{\star}{=}\mathbf{\hat{S}_{\star}}{\cdot}\mathbf{\hat{L}}$. 

For $\Omega_{\star}{<}\Omega$, Equations \ref{equation:tides} produce damping of the obliquity ($\dot{\psi_{\star}}{<}0$) in addition to planetary orbital decay and stellar spin evolution. To clarify how $\psi_{\star}$ will evolve under mass transfer and tides, we combine Equations \ref{equation:tides} and \ref{equation:mt} to obtain the following equation of motion for $\psi_{\star}$:

\begin{equation}\label{equation:dedt}
\hspace{-0.6cm}\frac{d\psi_{\star}}{dt}=
\begin{cases}
  -\frac{1}{t_{\rm tid}}\frac{L}{2S_{\star}}\sin\psi_{\star}\bigg[1-\frac{\Omega_{\star}}{2\Omega}(\cos\psi_{\star}-\frac{S_{\star}}{L})\bigg], & \dot{m}_{\rm RLO}=0\\
  -\beta\Gamma\frac{1}{t_{\rm RLO}}\frac{L}{S_{\star}}\sin\psi_{\star}, & \dot{m}_{\rm RLO}\neq0,
\end{cases}
\end{equation}

\noindent with 

\begin{align}\label{equation:ttid}
\begin{split}
    t_{\rm tid}&=\frac{Q_{\star}}{3k_{2,\star}}\frac{M_{\star}}{M_{\rm p}}\bigg(\frac{a}{R_{\star}}\bigg)^{5}\frac{1}{\Omega},\\
    &\sim 3.5 \bigg(\frac{Q_{\star}}{10^{5}}\bigg)\bigg(\frac{M_{\rm p}}{M_{\rm J}}\bigg)^{-1}\bigg(\frac{P_{\rm orb}}{1.45 \ \rm d}\bigg)^{13/3} \ \rm Gyr
\end{split}
\end{align}

\noindent the tidal timescale, and 

\begin{align}
\begin{split}
    t_{\rm RLO}&=\frac{M_{\rm p}}{|\dot{m}_{\rm RLO}|},\\
    & {\sim} 10^{4-5} \ \rm yr
\end{split}
\end{align}

\noindent the mass transfer timescale. Equation \ref{equation:dedt} reflects the fact that both tides and mass transfer drive $\psi_{\star}{\rightarrow}0$ (alignment), with tides dominating before/after RLO and mass transfer dominating whenever it is active. Whether tides or mass transfer ultimately drive $\psi_{\star}{\rightarrow}0$ depends on the angular momentum ratio $L/S_{\star}$ as well as the tidal and mass loss timescales. Tides will damp $\psi_{\star}$ during orbital inspiral ($\textit{before}$ mass transfer) if the tidal damping time $t_{\rm tid, \psi_{\star}}{\sim}2t_{\rm tid}S_{\star}/L$ is shorter than the orbital decay time $t_{\rm tid, a}{\sim}2t_{\rm tid}/13$ \citep[accounting for a factor 2/13 following][]{barogi09}. Setting $t_{\rm tid, \psi_{\star}}{\sim}t_{\rm tid,a}$, only $L/S_{\star}{\gtrsim}13$ can induce tidal damping of the obliquity before RLO. This angular momentum ratio translates to an initial stellar spin period of (Equation \ref{equation:LS}),

\begin{align}\label{equation:Pstarobl}
\begin{split}
    P_{\star,\rm crit}\sim 35 \bigg(\frac{M_{\rm p}}{M_{\rm J}}\bigg)^{-1}\bigg(\frac{M_{\star}}{M_{\odot}}\bigg)^{1/3}\bigg(\frac{R_{\star}}{R_{\odot}}\bigg)^{2}\bigg(\frac{P_{\rm orb}}{2 \ \rm d}\bigg)^{-1/3} \ \rm d.
\end{split}
\end{align}

\noindent Tides alone can damp obliquity for stars with initial rotation periods greater than this critical period, i.e. for $P_{\star}{\gtrsim}P_{\star, \rm crit}$. This threshold stellar rotation period is longer than 
that of main sequence FGK stars \citep[][]{vanceimat16}, so we expect
mass transfer to dominate obliquity damping unless the host star is exceptionally slowly rotating, or the hot Jupiter is particularly massive. All roads, however, lead to Rome: systems should tend toward alignment after RLO.

\begin{figure*}[t]
\centering
\includegraphics[width=1\textwidth]{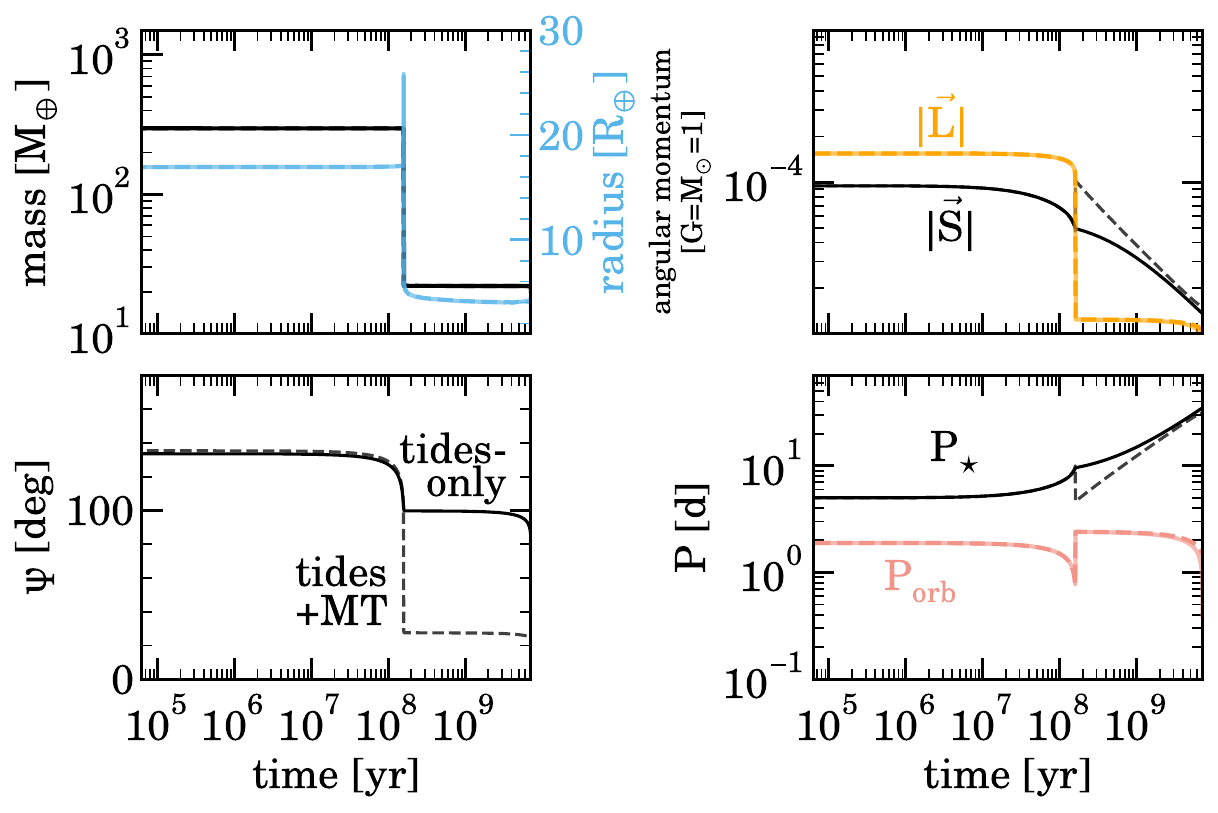}
\caption{Dynamical evolution of a star/hot Jupiter system through Roche lobe overflow. From top to bottom, left panels depict: planet mass (left axis; black) and radius (right axis; blue), and obliquity. From top to bottom, right panels show: angular momenta in natural units (orbital angular momentum in orange, stellar spin angular momentum in black), and period (orbital in salmon, stellar spin in black).
Solid curves assume that only tides transfer orbital angular momentum to the star (${\beta}{=}0$; labeled ``tides-only"), while dashed curves include angular momentum transfer during RLO (${\beta}{=}1$; ``tides+MT"). Assuming angular momentum transfer during RLO is modulated by the Alfvén radius ($\beta(R_{\rm A})$) yields an intermediate obliquity between the tides-only and tides+MT curves. 
This example employs $Q_{\star}{=}10^{4}$, initial planet entropy $S{=}9.5 \ k_{\rm B}/m_{\rm H}$, and a solar composition planetary atmosphere. Planet structure/mass loss calculations follow \cite{halmil26}. Stellar obliquity is damped below ${\lesssim}25^{\circ}$ by mass transfer (for ${\beta}{=}1$, or ${<}45^{\circ}$ when using $\beta(R_{\rm A})$), not tides (${\beta}{=}0$), when initial stellar spin and orbital angular momenta are commensurate.
\label{figure:evo_example}}
\end{figure*}

\subsubsection{Magnetic Braking}

Host stars spun up via tides and mass transfer will shed spin angular momentum via magnetic braking \cite[e.g.][]{webdav67,matmacpin12}. We track magnetic spin down following the standard Skumanich braking law whereby the stellar rotation velocity ${\propto}t^{-1/2}$ \citep[][]{sku72}:

\begin{align}
\begin{split}
    \frac{d\mathbf{S_{\star}}}{dt}\bigg|_{\rm MB}&=\alpha_{\star}M_{\star}R^{2}_{\star}\frac{d\Omega_{\star}}{dt}\bigg|_{\rm MB}\mathbf{\hat{S}_{\star}},\\
    \frac{d\Omega_{\star}}{dt}\bigg|_{\rm MB}&=-\phi_{\rm MB}\Omega_{\star}\texttt{min}(\Omega_{\star},\Omega_{\rm sat})^{2}
\end{split}
\end{align}

\noindent where $\phi_{\rm MB}{=}1.5{\times}10^{-14}$ yr \citep[][]{doblinmar04}, and we have accounted for saturation of the braking torque at high rotation rates with $\Omega_{\rm sat}{=}10\Omega_{\odot}$ where $\Omega_{\odot}{=}2\pi/(25.67 \ \rm days)$ \citep[e.g.][]{silpinter00,galbou13,elbconful22}.

\subsection{Summary of Calculation}

In total, we integrate the following equations of motion for orbital and stellar angular momenta:

\begin{align}
\begin{split}
    \frac{d\bf{L}}{dt}&=\frac{d\bf{L}}{dt}\bigg|_{\rm MT}+\frac{d\bf{L}}{dt}\bigg|_{\rm PE}+\frac{d\bf{L}}{dt}\bigg|_{\rm tid},\\
    \frac{d\bf{S}_{\star}}{dt}&=\frac{d\bf{S}_{\star}}{dt}\bigg|_{\rm MT}+\frac{d\bf{S}_{\star}}{dt}\bigg|_{\rm MB}+\frac{d\bf{S}_{\star}}{dt}\bigg|_{\rm tid}.\\
\end{split}
\end{align}

\noindent Planetary mass loss/structure calculations follow \cite{halmil26}. We explicitly confirm in the \hyperref[sec:appendix]{Appendix} that the \cite{halmil26,halmil26b} method to couple planet structure/mass loss/orbital evolution is accurate, producing similar results as full-fledged hydrodynamics simulations with $\texttt{MESA}$ \citep[][]{paxbildot11,paxcanarr13,paxmarsch15,paxschbau18,paxsmosch19}. We perform Monte Carlo simulations varying the initial (pre-inspiral) stellar obliquity uniformly in $\cos\psi_{\star}{\in}[-1,1]$. Each Monte Carlo ensemble consists of 300 integrations (300 randomly sampled values of $\cos\psi_{\star}$), using planet/star initial conditions following Table \ref{tab:fiducial_table}. Integrations run from 1 to 10 Gyr, or stop once remnant planets spiral into the stellar surface. We employ $\texttt{scipy}$'s implicit $\texttt{BDF}$  integrator \citep[][]{virgomoli20} with relative and absolute tolerances of $10^{-12}$. In rare cases when we find that the integration fails during peak mass transfer, we re-run with relaxed tolerances of $10^{-7}$.

\section{Tilting Stars Into Alignment During RLO}\label{sec:obl}

\begin{figure*}[t]
\centering
\includegraphics[width=1\textwidth]{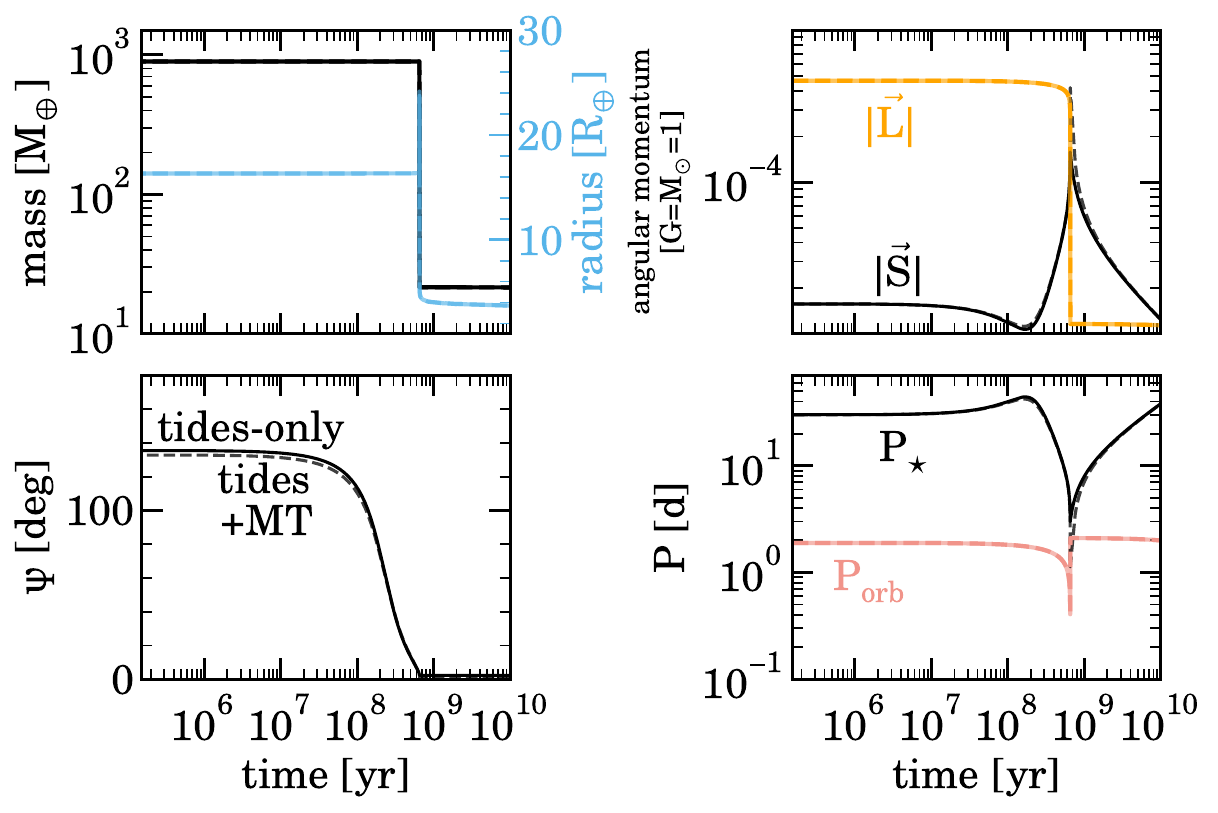}
\caption{Similar to Figure \ref{figure:evo_example}, except with initial angular momentum budget dominated by the planet orbit. To explore this $L{\gg}S_{\star}$ regime, this example uses a $900 \ M_{\oplus}$ hot Jupiter as well as a slowly rotating star ($P_{\star}{=}$30 days; we also use a stellar tidal quality factor $Q_{\star}{=}10^{5}$). When orbital angular momentum dominates the stellar spin, obliquity tides damp the system into spin/orbit alignment $\textit{before}$ RLO (according to our weak friction, equilibrium tide approach).
\label{figure:evo_example2}}
\end{figure*}

\begin{figure}
\epsscale{1.15}
\plotone{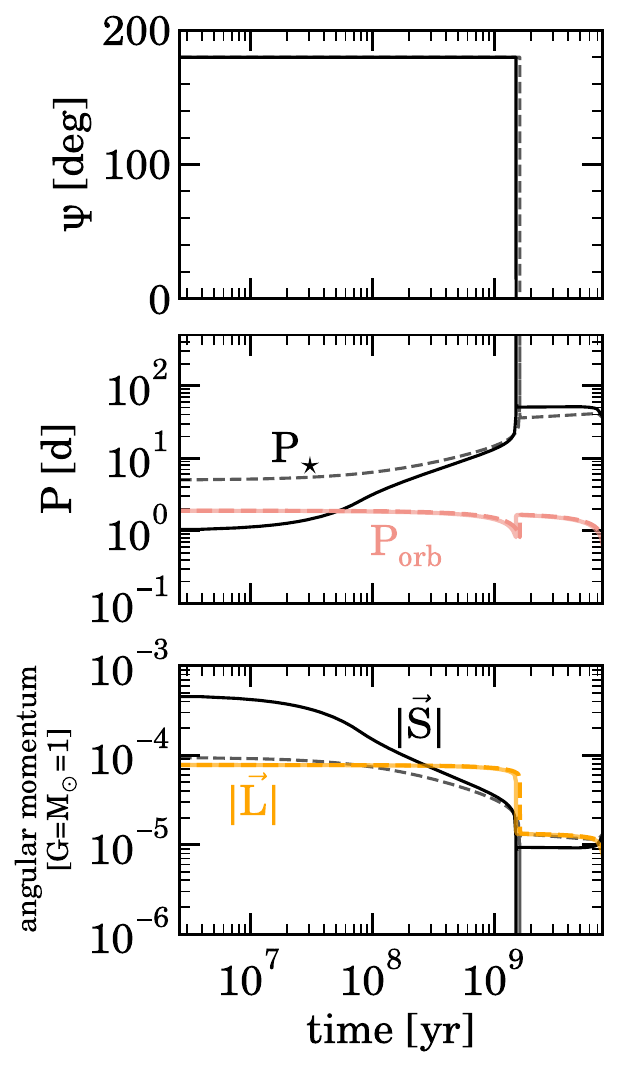}
\caption{Evolution of fully retrograde systems (initial $\psi_{\star}{=}180^{\circ}$) during RLO. Two examples are displayed: a very fast rotator (solid lines, with $P_{\star}{=}1$ day; $L{<}S_{\star}$) and moderately fast (dashed lines, $P_{\star}{=}5$ days; $L{\sim}S_{\star}$). This example uses an initial planet mass and entropy 150 $M_{\oplus}$ and $S{=}8 \ k_{\rm B}/m_{\rm H}$ respectively, $Q_{\star}{=}5{\times}10^{4}$, and employs a variable angular momentum transfer fraction, $\beta(R_{\rm A})$. Stellar spin down during orbital inspiral drives the system into alignment; global maxima/minima in $P_{\star}$ and $S_{\star}$ respectively correspond to reversal in the direction of $\bf{S}_{\star}$ ($P_{\star}{\rightarrow}\infty$).  Stars can rotate slowly after RLO, if systems begin strongly retrograde. \label{figure:retrograde}}
\end{figure}

\subsection{Mass Transfer vs. Tides}

Before showcasing the population-level obliquity trends predicted by RLO, we first consider two example systems that underscore how mass transfer and stellar tides dictate a system's final obliquity.

Figure \ref{figure:evo_example} displays a star/planet system with initially commensurate spin/orbit angular momenta ($S_{\star}{\sim}L$). We chose a fast initial stellar rotation period below the 
critical period required for tides to damp obliquity (Equation \ref{equation:Pstarobl}) to highlight the efficacy of mass transfer. Isolating the effect of tides on obliquity damping (by setting $\beta{=}0$; Equation \ref{equation:mt}) confirms that tidal dissipation alone cannot align the stellar spin and planet orbit. The system can be brought into near alignment, with final obliquity ${\psi_{\star}}{\sim}25^{\circ}$, only due to the flow of angular momentum due to mass transfer ($\beta{=}1$). While not pictured in Figure \ref{figure:evo_example}, if we account for angular momentum transfer modulated by the Alfvén radius using $\beta(R_{\rm A}(t))$, we find intermediate obliquity damping with final obliquity ${\psi_{\star}}{\sim}45^{\circ}$. Using $\beta$ set by the Alfvén radius produces intermediate obliquity damping because, at peak RLO \citep[when $|\dot{m}_{\rm RLO}|{\sim}10^{9} \ M_{\oplus}$/Gyr;][]{halmil26}, outflowing gas' ram pressure exceeds magnetic pressure down to the stellar surface, pushing $R_{\rm A}{\rightarrow}R_{\odot}$ (Equation \ref{equation:ralf}). At the height of mass transfer, the angular momentum accreted by the star is therefore limited to that at $R_{\odot}$. 

In contrast, Figure \ref{figure:evo_example2} displays a system with angular momentum budget dominated by the planet orbit. To shift the angular momentum budget to the planet orbit, we employ a $3{\times}$ more massive hot Jupiter than in Figure \ref{figure:evo_example} as well as a $6{\times}$ larger initial stellar rotation period. In this $L{\gg}S_{\star}$ case, tides damp obliquity ($\psi_{\star}{\rightarrow}0^{\circ}$) rapidly during pre-RLO inspiral. Mass transfer does not affect the already damped obliquity. Figures \ref{figure:evo_example} and \ref{figure:evo_example2} underscore that, regardless of whether tides or mass transfer dominate obliquity evolution, the result is the same: the stellar obliquity is shepherded into alignment.

Obliquity damping is even more pronounced in fully retrograde systems. Figure \ref{figure:retrograde} showcases an example wherein orbit and stellar spin are initially anti-aligned. Tidal spin-down during orbital inspiral switches obliquity into full alignment before mass transfer begins ($\vec{S}_{\star}$ reverses sign, passing through $P_{\star}{=}\infty$), even when $S_{\star}{>}L$ initially. In contrast to systems that begin more aligned, Figure \ref{figure:retrograde} further highlights that initially anti-aligned systems can produce very slowly rotating stars after mass transfer. Slow stellar rotation after mass transfer may alleviate tension between RLO and the well-known desert dweller LTT 9779 b \citep[][]{jendiamat20}. As pointed out by \cite{halmil26}, LTT 9779 b's present-day mass, radius and orbital period are readily reproduced by RLO \citep[aided by weak photoevaporation due to low stellar X-ray emission;][]{ferwhekin24}, but its host star's slow rotation rate \citep[${\sim}45/\sin i$ days;][]{jendiamat20} runs counter to predicted tidal spin up in ``standard" RLO. 

While most hot Jupiters around GK stars are near alignment \citep[e.g.][]{albdawwin22}, at least two anti-aligned hot Jupiters or proto-hot Jupiters \textit{do} exist \citep[with projected obliquity ${\lambda_{\star}}{\sim}180^{\circ}$ for KELT-23A b and stellar obliquity $\psi_{\star}{=}141^{+15^\circ}_{-24}$ for TIC 241249530 b respectively;][]{gupmilim24,giahowrub25}. We conclude that RLO remains a viable explanation for LTT 9779 b.

Given that the Alfvén radius provides a lower limit on $\beta{\gtrsim}0.5$, we expect that significant angular momentum should always be transferred to the star during RLO; $\beta{=}0$ is unrealistically low. We therefore conclude that regardless of the initial angular momentum distribution, mass transfer and tides ensure that obliquities are damped after RLO. The next section confirms that alignment within ${\sim}$a few tens of degrees is the default outcome of RLO across parameter space.

\subsection{Population Study}

\begin{figure*}
\centering
\includegraphics[width=1\textwidth]{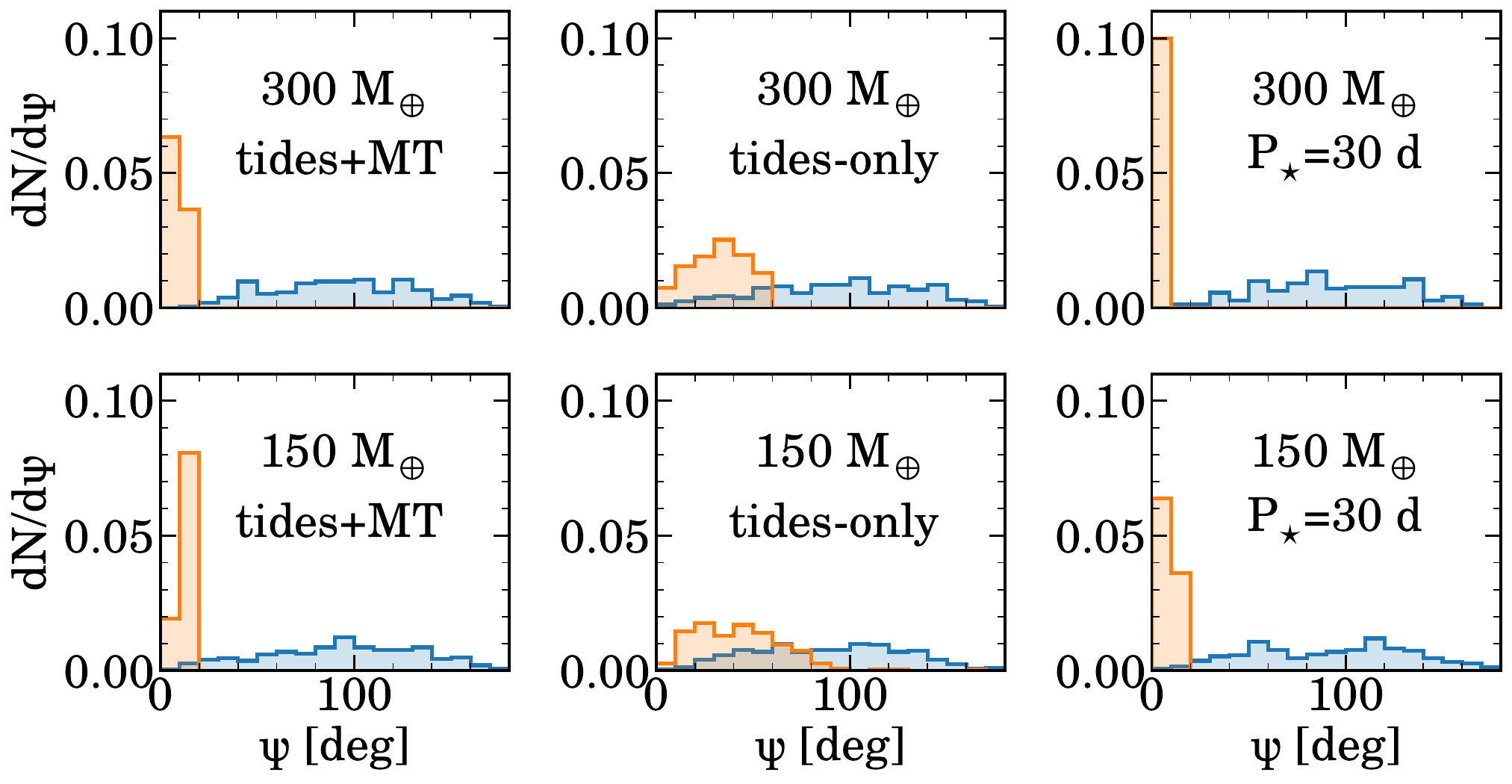}
\caption{Stellar obliquity ($\psi_{\star}$) distributions sculpted by RLO. Blue histograms ($\cos \psi_{\star}$ uniformly sampled in [-1,1]) are processed by RLO into the orange distributions. We adjust the initial angular momenta budget for each ensemble by varying initial planet mass and stellar rotation period. Leftmost panels use an initial stellar rotation period $P_{\star}{=}5$ days, and assume perfect transfer of angular momentum during mass transfer (${\beta}{=}1$; ``tides+MT"). Middle panels isolate the effect of tidal alignment during orbital inspiral by ignoring angular momentum transfer due to RLO (${\beta}{=}0$). Rightmost panels are the same as the lefmost, except they use an initially slowly rotating host star with $P_{\star}{=}$30 days. Tides and angular momentum transfer ensure that all roads lead to Rome; RLO robustly predicts that stellar spins are aligned within ${\sim}$a few tens of degrees with the orbits of remnant planets.
\label{figure:hists}}
\end{figure*}

In Figure \ref{figure:hists} we display synthetic obliquity distributions from our Monte Carlo ensembles. In all cases, the wide initial obliquity distribution collapses into a strongly aligned distribution. As expected, cases with zero angular momentum transfer during RLO exhibit a wider final distribution than cases with angular momentum transfer included, although maximum final obliquities are all less than ${\lesssim}60^{\circ}$. Changes in the initial angular momentum ratio (e.g. by adjusting the initial hot Jupiter mass or stellar rotation period) produce minor deviations in the final obliquity distributions; even in the $S_{\star}{\gtrsim}L$ case (150 $M_{\oplus}$ hot Jupiter around a rapidly rotating star; top right panel) in which tidal alignment is minimized (maximum $\psi_{\star}{\sim}80^{\circ}$ in the tides-only case), mass transfer supplies sufficient angular momentum to produce very strong obliquity damping. The lesson of Figure \ref{figure:hists} is that mass transfer robustly predicts damped stellar obliquities.

In Section \ref{sec:companion}, we consider how deviations from the prediction of spin/orbit alignment may arise during post-RLO system evolution.

\vspace{0.5cm}

\section{Post-RLO Evolution}\label{sec:companion}

We have demonstrated that RLO damps stellar obliquities through tidal dissipation and angular momentum transfer. While damped stellar obliquities are a clean prediction of the \cite{halmil26} mechanism, in this section we consider how deviations from predicted alignment can arise after RLO due to external planets in the system. We estimate companion mass/orbital distance/inclination under which desert dwellers undergo obliquity excitation, to guide future observations testing the \cite{halmil26} picture.

We consider the effect of a companion because the additional torque on a desert dweller's orbit can excite obliquity by inclining the desert dweller, rather than tilting the star \citep[e.g.][]{han17}. Orbit tilting need not be resonant; obliquity excitation can occur as the dominant angular momentum vector evolves from that of the star (shortly after RLO) to that of the outer companion as the star spins down \citep[e.g.][]{spamil20}. In the absence of a companion, a desert dweller's orbit normal vector and host star spin vector mutually precess about their total angular momentum vector $\bf{S}_{\star}{+}\bf{L}$ at fixed obliquity \citep[e.g.][]{harwar82,tre23}. Adding a third body of mass $M_{\rm C}$ and semi-major axis $a_{\rm C}$, the relevant nodal precession frequencies read \citep[e.g.][]{matkon17}, 

\begin{align}\label{equation:omprec}
\begin{split}
    \omega_{\rm \star\rm P}&=\frac{3J_{2}}{2\alpha_{\star}}\frac{GM_{\rm p}}{a^{3}\Omega_{\star}},\\
    \omega_{\rm PC}&=\frac{3M_{\rm C}}{4M_{\star}}\bigg(\frac{a}{a_{\rm C}}\bigg)^{3}\Omega,\\
\end{split}
\end{align}

\noindent with $J_{2}{=}k_{2,\star}\Omega^{2}_{\star}R^{3}_{\star}/(3GM_{\star})$ the stellar quadrupole moment \citep[][]{murder99}. The frequency $\omega_{\rm \star\rm P}$ corresponds to star/inner planet nodal precession, whereas the frequency $\omega_{\rm PC}$ arises due to nodal precession of the inner planet due to the outer (the star and outer planet also mutually precess, but those frequencies are not relevant to our estimates here).

Obliquity can only be excited if torque from the external companion disrupts the star/inner planet coupling. The torque from a companion dominates when the star/inner planet precession frequency, $\omega_{\star \rm P}$, equals that of the inner planet/companion, $\omega_{\rm PC}$. 
Obliquity can be pumped by a companion at orbital distance,

\begin{align}\label{equation:res}
\begin{split}
    a_{\rm C}&=\bigg(\frac{3\alpha_{\star}}{2k_{2,\star}}\frac{M_{\rm C}}{M_{\rm p}}\frac{\Omega}{\Omega_{\star}}\bigg)^{1/3}\frac{a^{2}}{R_{\star}},\\
    &\hspace{-1cm}\sim1.7\bigg(\frac{P_{\rm orb}}{2 \ \rm d}\bigg)\bigg(\frac{P_{\star}}{20 \ \rm d}\bigg)^{1/3}\bigg(\frac{M_{\rm C}}{M_{\rm J}}\bigg)^{1/3}\bigg(\frac{M_{\rm p}}{20 \ M_{\oplus}}\bigg)^{-1/3} \ \rm au.
\end{split}
\end{align}

\begin{figure}
\epsscale{1.15}
\plotone{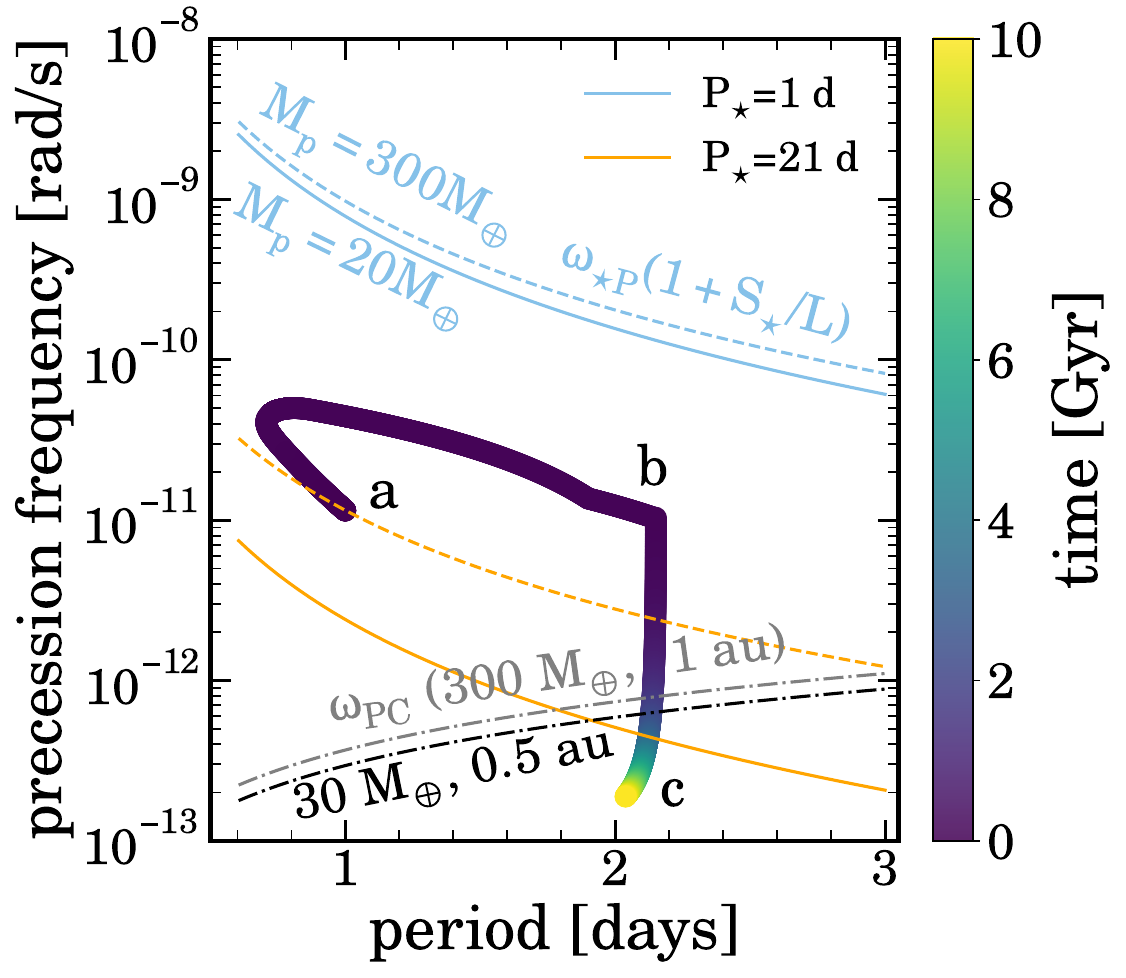}
\caption{Precession frequency evolution versus orbital period following mass transfer. Following \cite{laiandpu18}, when torque from an external planet disrupts star/inner planet precession ($\omega_{\star \rm P}(1{+}S_{\star}/L){\sim}\omega_{\rm PC}$), the inner planet's obliquity may be excited. Blue/orange curves depict $\omega_{\star \rm P}(1{+}S_{\star}/L)$ for fast/slow rotating stars (see legend), while dashed/solid curves correspond to 300 and 20 $M_{\oplus}$ inner planets, respectively. The grey/black dot-dashed curves denote the planet/planet precession frequency for differing external companion parameters (labelled). The colored curve, color bar, and points a, b, c depict an example of $\omega_{\star \rm P}(t)(1{+}S_{\star}(t)/L(t))$ evolution during mass transfer; the example planet begins at period ${\sim}1$ day (point a) before bouncing out to ${\sim}$2 days after RLO (point b), after which the star spins down via magnetic braking (point c). Stellar spin down permits orbit tilting when $\omega_{\star \rm P}(1{+}S_{\star}/L){\sim}\omega_{\rm PC}{\sim}8{\times}10^{-13}$ rad/s (between b and c). Hot Jupiter remnants' obliquities can be excited by (as yet unseen) companion planets. \label{figure:prec}}
\end{figure}

\noindent We therefore expect that companions to misaligned desert dwellers should exist within ${\lesssim}$2 au. Such companions need not be massive; even a $5 \ M_{\oplus}$ perturber can excite obliquity at $a_{\rm C}{\sim}0.4$ au. That distant companions can excite stellar obliquities differs from the case of ultra-hot Jupiters, whose orbits cannot easily be tilted by companions.\footnote{We note that our simplified estimate for when orbit tilting occurs, $\omega_{\star \rm P}{\sim}\omega_{\rm PC}$, differs from that of \cite{laiandpu18} by a factor $(1{+}S_{\star}/L)$ (see their Equation 9), which we find to be ${\sim}2$ for the post-RLO epoch when $S_{\star}$ has braked to ${\sim}L$. This difference in resonance estimate changes $a_{\rm C, \rm res}$ by a factor $2^{1/3}{\sim}1$.} 
Figure \ref{figure:prec} displays an example evolution track during which an outer companion can pump obliquity.

The perturber's inclination required to pump obliquity depends on whether the excitation is resonant. A secular spin/orbit resonance is crossed once $\omega_{\star \rm P}{\sim}\omega_{\rm PC}$, provided $S_{\star}{<}L$ \citep[][]{laiandpu18}.\footnote{Obliquity excitation driven by secular spin/orbit resonance when $\omega_{\star\rm P}{\sim}\omega_{\rm PC}$ is distinct from secular resonance induced by the stellar quadrupole in systems of close-in planets, where inner/outer planets precess at similar rates \citep[e.g.][]{spabat16,farnaoli23}.}
Outside resonance, exciting obliquity $\psi_{\star}$ requires a perturber to be inclined by ${\sim}\psi_{\star}$. Mass transfer usually enforces $S_{\star}{>}L$ for Gyr after RLO (see e.g. Figure \ref{figure:evo_example}), so that obliquity excitation would be non-resonant for most desert dwellers. Resonance can, however, occur for desert dwellers ${\gtrsim}30 \ M_{\oplus}$ (to enable $L{>}S_{\star}$ at the time precession frequencies cross); such massive desert dwellers can be tilted even if their companion is not initially strongly misaligned. 

As a concrete example, the TOI-1288 system \citep[$44.1^{+2.7}_{-2.6} \ M_{\oplus}$ at 2.7 days, companion $85.7 / \sin i \  M_{\oplus}$ at 1 au;][]{pollubbea24,knugannow23} can traverse this resonant regime. TOI-1288's considerable distance \citep[${\sim}115$ pc;][]{knugannow23} yields a $\textit{Gaia}$ astrometric signal to noise ${\sim}0.1$ \citep[using $\textit{Gaia}$ single-scan uncertainty 0.034 marcsec., and astrometric signature ${\sim}2{\times}10^{-6}$ arcsec.;][]{ranhoblin18}, well below the detection threshold for Data Release 5; dedicated system characterization would therefore help determine whether RLO remains a viable explanation for TOI-1288 b. 

If follow-up observations of desert dwellers confirm the presence of external companions, our work may constrain the origin channel of their progenitor hot Jupiters. Depending on companion parameters (mass, eccentricity, inclination, etc.) they may be consistent with a desert dweller origin via secular chaos \citep[][]{wulit11} or coplanar HEM \citep[][]{pet15b}, while pointing against von-Zeipel/Lidov/Kozai (ZLK) oscillations \citep[e.g.][]{wumur03,fabtre07} if the companion parameters do not permit planet/planet ZLK excitation. 

\begin{deluxetable*}{ccccc}\label{tab:expectations}
\tablecaption{Population-Level Expectations From Lossy RLO vs. HEM.} 
\label{tab:parameters}
\tablecolumns{4}
\tablewidth{0pt}
\tablehead{
\colhead{Attribute} &
\colhead{Lossy RLO} &
\colhead{HEM} & 
\colhead{Reference} & 
}
\startdata
\rm stellar \ obliquities & \parbox[c]{5.5cm}{\centering\vspace{4pt}${\lesssim}20^{\circ}(\beta{=}1)$,\\ ${\lesssim}60^{\circ}(\beta{=}0)$} & \parbox[c]{5.5cm}{\centering\vspace{4pt} ${\lesssim}160^{\circ}$ ($\rm stellar \ ZLK$), \\ ${\lesssim}130^{\circ}$ ($\rm planet \ ZLK$), \\ ${\lesssim}90^{\circ}$ ($\rm chaos$), \\ ${\lesssim}45^{\circ}$ ($\rm coplanar$) \vspace{4pt}} & \parbox[c]{1.2cm}{\centering\vspace{3pt}(1)\\(2)\\(3)\\(4)\vspace{3pt}}  \\
\rm stellar \ age & ${\sim}$Gyrs & ${\gtrsim}$0.1 \ \rm Gyr & (1,2,3) \\
\rm stellar \ type & \rm FGK, \ below \ Kraft \ break  & \rm FGK & (5) \\
\rm nearby \ companion  \ planets  & \rm compatible & \rm incompatible & (6,7) \\
\rm desert \ dweller \ density & $1{-}10$ \ \rm g/cm$^{3}$ & \rm undetermined &  \\
\rm distant \ companions \ (planet \ or \ star) & \rm same \ as \ hot \ Jupiters & \rm same \ as \ hot \ Jupiters &  \\
\enddata
\tablerefs{ (1) - \cite{andstolai16}, (2) - \cite{pettre16}, (3) - \cite{teylaivic19}, (4) - \cite{pet15b}, (5) - \cite{dawjoh18}, (6) - \cite{livhalmil26}, (7) - \cite{musdavjoh15}. }
\end{deluxetable*}

\section{Distinguishing RLO From HEM}\label{sec:HEM}

In this section, our goal is to clarify how desert dweller formation via lossy RLO can be observationally distinguished from tidal disruption during HEM. Readers interested in a summary of divergent predictions between RLO and HEM may consult Table \ref{tab:expectations} and Figure \ref{figure:population_schematic}. The rest of this Section elaborates on the predictions included in Table \ref{tab:expectations}. We first consider post-HEM tidal realignment, finding that it is inefficient and thus preserves the distinct obliquity trends predicted by RLO and HEM. Stellar type dependencies and the presence of nearby companion planets are then discussed. We close with evaluation of observational prospects.

\subsection{Post-HEM Tidal Realignment}

Desert dwellers formed via lossy RLO $\textit{must}$ exhibit low stellar obliquities if they do not have misaligned companion planets. Here we quantify how this spin/orbit alignment differentiates RLO from the obliquity distribution predicted by HEM. At first glance, it is plausible that the broad obliquity distribution produced by HEM may be post-processed via stellar tides into alignment, as suggested for hot Jupiters around cool stars \cite[e.g.][]{winfabalb10,albdawwin22}. Such post-migration realignment would prevent differentiating RLO from HEM via their obliquity distributions. We will show however that, due to their lower masses, desert dwellers likely avoid such tidal damping, preserving the uniqueness of RLO's predicted spin/orbit alignment.

Following our weak friction tidal formulation in Section \ref{subsubsec:tides}, the timescale to realign a desert dweller's orbit with the stellar spin after HEM is given by $t_{\rm tid,\psi_{\star}}{\sim}2t_{\rm tid}S_{\star}/L$, with $t_{\rm tid}$ the tidal timescale (Equation \ref{equation:ttid}). Inserting the angular momenta (Equation \ref{equation:LS}), the obliquity damping time reads,

\begin{align}\label{equation:damping}
\begin{split}
    t_{\rm tid,\psi_{\star}}&= \frac{1}{12 \pi^{3}}\frac{\alpha_{\star}Q_{\star}}{k_{2,\star}}\bigg(\frac{M_{\star}}{M_{\rm p}}\bigg)^{2}\omega^{2}_{\star,\rm dyn}\frac{P^{4}_{\rm orb}}{P_{\star}} \\
    &\approx 30 \bigg(\frac{Q_{\star}}{10^{5}}\bigg)\bigg(\frac{M_{\rm p}}{0.1 \ M_{\rm J}}\bigg)^{-2}\bigg(\frac{M_{\star}}{M_{\odot}}\bigg)^{3}\bigg(\frac{R_{\star}}{R_{\odot}}\bigg)^{-3} \\ 
    & \times \bigg(\frac{\alpha_{\star}}{0.06}\bigg)\bigg(\frac{P_{\rm orb}}{1  \ \rm d}\bigg)^{4}\bigg(\frac{P_{\star}}{20 \ {\rm d}}\bigg)^{-1} \ \rm Gyr 
\end{split}
\end{align}

\noindent with $\omega_{\star,\rm dyn}{=}\sqrt{GM_{\star}/R^{3}_{\star}}$ the stellar dynamical frequency. Equation \ref{equation:damping} shows that while Jupiter-mass planets can realign quickly in just ${\sim}0.3$ Gyr, Neptunian planets in the desert are unlikely to tidally realign. Neptunian planets can only realign within ${\lesssim}$10 Gyr if they are extremely close to their stars, with periods less than the critical period $P_{\rm orb,crit}$ given by:

\begin{figure*}[t]
\centering
\includegraphics[width=0.8\textwidth]{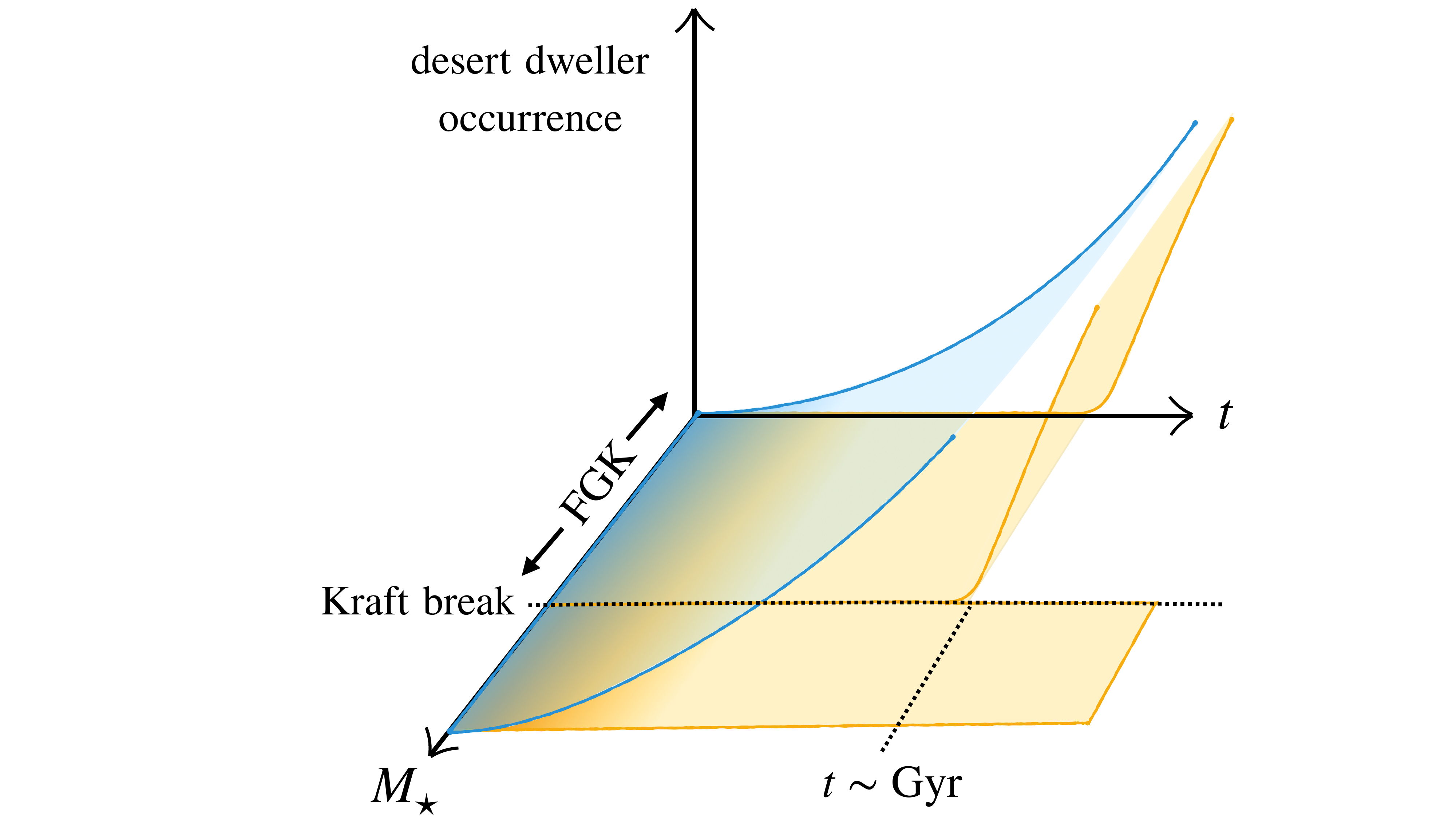}
\caption{Schematic of population-level expectations for desert dweller occurrence (vertical axis) versus stellar properties (stellar mass $M_{\star}$ and age $t$ in $xy$ plane), in the lossy RLO (orange) and high eccentricity migration (blue) pictures. In contrast to HEM, lossy RLO predicts that desert dwellers are emplaced over ${\sim}$Gyr due to long-term orbital decay from stellar tides. Lossy RLO also predicts that the sub-Jovian desert should remain empty around stars above the Kraft break due to weaker stellar tides, while HEM should operate indiscriminately. In addition to obliquities, desert dwellers' occurrence with stellar properties can distinguish their origin channels.}
\label{figure:population_schematic}
\end{figure*}

\begin{align}\label{equation:Porbdamp}
\begin{split}
    P_{\rm orb,\rm crit}&\sim0.75\bigg(\frac{Q_{\star}}{10^{5}}\bigg)^{-1/4}\bigg(\frac{M_{\rm p}}{0.1 \ M_{\rm J}}\bigg)^{1/2}\bigg(\frac{M_{\star}}{M_{\odot}}\bigg)^{-3/4}\\
    &\hspace{-1.15cm}\times \bigg(\frac{\alpha_{\star}}{0.06}\bigg)^{-1/4}\bigg(\frac{R_{\star}}{R_{\odot}}\bigg)^{3/4} \bigg(\frac{P_{\star}}{20 \ \rm d}\bigg)^{1/4}\bigg(\frac{t_{\rm tid,\psi_{\star}}}{10 \ \rm Gyr}\bigg)^{1/4} \ \rm d.
\end{split}
\end{align}

\noindent We therefore expect that any desert dweller emplaced by HEM beyond $P_{\rm orb,crit}{\gtrsim}0.75$ days (${\gtrsim}1.3$ days for $Q_{\star}{=}10^{4}$) cannot realign afterward. In other words: beyond $P_{\rm orb,crit}$, alignment alone differentiates RLO from HEM. Observations should focus on planets beyond $P_{\rm orb,crit}$ to cleanly compare RLO and HEM.

Our conclusion that desert dwellers cannot tidally realign is bolstered by empirical constraints on the reach of stellar tides. Observations suggest that hot Jupiters around cool stars beyond orbital separations $a/R_{\star}{\gtrsim}8$ (orbital period ${\sim}2.6$ days for a Sun-like star) cannot realign \citep[see Figure 10 of][]{albdawwin22}. Given that the tidal realignment timescale goes as ${\propto}(M_{\star}/M_{\rm p})^{-2}$, the corresponding critical period for desert dweller realignment is ${\lesssim}0.8$ days, in line with our estimate in Equation \ref{equation:Porbdamp}. Alternative tidal formulations, e.g. inertial wave components that damp obliquities but not semi-major axes \citep[][]{lai12} or a low effective $\alpha_{\star}$ from decoupled stellar layers \citep[see][for discussion and caveats of this approach]{daw14}, are unlikely to change our qualitative conclusion. In general, the smaller mass ratio in desert dweller systems makes stellar tides less efficient compared to hot Jupiters. As an example, setting $\alpha_{\star}{=}0.0008$ following \cite{albdawwin22} (assuming such a small value is realistic) we find $P_{\rm orb,crit}{\sim}2 \ \rm d$, so HEM and RLO should still produce different predictions across the desert \citep[defined ${\lesssim}$3.2 days;][]{casboulil24}.

\subsection{Population-Level Imprints of ``Lossy" RLO}\label{subsection:signatures}

Having estimated the obliquity distribution arising from lossy RLO, and quantified how this obliquity prediction differs from that of HEM, Table $\ref{tab:expectations}$ summarizes the population-level expectations each picture entails. Figure \ref{figure:population_schematic} also provides a visual summary of the population-level trends with stellar properties. The desert dweller obliquity distribution predicted by HEM depends on the flavor of HEM (coplanar, stellar ZLK, etc.), but higher obliquities are expected than those produced by lossy RLO. HEM also predicts that hot planets can be emplaced early, in contrast to tidal decay-driven RLO which 
necessarily requires ${\sim}$Gyr \citep{halmil26}.

Another avenue to test the \cite{halmil26} theory is by characterizing the sub-Jovian desert around stars with effective temperatures above the Kraft break \citep[${\sim}6200{-}6550$ K;][]{kra67,beywhi24}. Stars hotter than the Kraft break are thought to produce weaker tidal dissipation owing to thinner convective envelopes and weaker magnetic braking than cooler stars \citep[e.g.][]{winfabalb10,albdawwin22}, potentially explaining the elevated stellar obliquities of hot stars with hot Jupiters. Mirroring this trend of weaker tidal interaction around hot stars, mass transfer driven by orbital inspiral is expected to become less frequent above the Kraft break; the sub-Jovian desert should remain empty. This expectation of an empty desert above the Kraft break contrasts with the expectation from hot Jupiter destruction during HEM, where tidal disruption should occur indiscriminately above/below the break.

The presence of nearby companion planets provides yet another line of distinction. HEM is known to disintegrate inner planetary systems \citep{musdavjoh15}, which contrasts with the ${\sim}10{-}20\%$ of hot Jupiters that $\textit{do}$ host neighboring planets \citep[see][who found a companion fraction $12{\pm}6\%$ accounting for interior and exterior nearby perturbers, and \citealt{shavanhua26} who report an interior companion fraction $7.6^{+5.5}_{-3.8}\%$]{wuricwan23}. On the other hand, \cite{livhalmil26} demonstrate that lossy RLO remains compatible with multiplanet systems containing a desert dweller (outer planets are stable during RLO of an inner hot Jupiter). Finally, we highlight that although calculations suggest tidally disrupted products of HEM can take a range of densities \citep[][]{guiramlin11,liuguilin13}, quantitative predictions for desert dwellers have yet to be made from the HEM picture. In contrast, \cite{halmil26} and the \hyperref[sec:appendix]{Appendix} of this work establish that lossy RLO can successfully reproduce desert dwellers' range of densities, ${\sim}1{-}10$ g/cm$^{3}$.

In sum, we expect that lossy RLO and HEM can be successfully discerned at the population level following Table \ref{tab:expectations} and Figure \ref{figure:population_schematic}. Determining whether RLO occurred in an individual system will, however, require bespoke modeling to reproduce the system in the present day (e.g. planet structure and orbit, stellar age and rotation, etc.), which is beyond the scope of this paper.

\subsection{Observational Prospects}

We close this section with brief discussion of the observational prospects for following up Table \ref{tab:expectations}. Desert dweller obliquities appear readily measurable as they are akin to ongoing efforts for planets surrounding the desert, following e.g. the $\texttt{DREAM}$ \cite[][]{bouattmall23,attia2023}, $\texttt{ATREIDES}$ \cite[][which already targets a few planets in the desert]{boustecas25}, and $\texttt{POSEIDON}$ \cite[][]{epswinbra26} surveys. Future completeness-corrected estimates of planet occurrence in the desert from $\textit{TESS}$ \citep{cuiarmhad26} above/below the Kraft break can test whether occurrence is lower around hotter stars. Forthcoming transiting planet demographic data from $\textit{Roman}$ \citep[][]{wilbarpow23} and $\textit{PLATO}$ \citep[][]{rauaercab25} may also sharpen measurements of desert dweller occurrence with stellar type and age. As $\textit{TESS}$ continues to grow the desert dweller sample \citep[e.g.][]{carcaspal26}, population-level age constraints based on stellar kinematics from $\textit{Gaia}$ will improve \citep[][]{halmil26}. Surveys like $\texttt{ATREIDES}$ and $\texttt{POSEIDON}$, in tandem with follow-up characterization of $\textit{TESS}$ targets, should also provide constraints on desert dweller age and stellar type distributions, as well as the presence of planetary companions.

\section{Discussion}\label{sec:discussion}

\subsection{Alternative Formation Channels}\label{subsec:alternatives}

Our paper has focused on how desert dweller formation via lossy RLO can be distinguished from HEM. One remaining possibility however is that formation/migration processes alternative to RLO/HEM are responsible for populating the sub-Jovian desert. If alternative formation routes are at play, spin/orbit alignment may not be a unique signature of RLO. Two alternative formation routes that also produce spin/orbit alignment are coplanar HEM \citep[CHEM;][]{linaokoc14,pet15b} and disk migration \citep[in disks aligned with stellar spin; e.g.][]{goltre79,goltre80,linpap79}. We address each of these alternative pathways in turn.

While CHEM can indeed shrink periapse distances below the tidal disruption limit, \cite{pet15b} found that such destruction is rare and only occurred as a result of extreme initial conditions instead of secular evolution; disruption occurred in ${\sim}0.1\%$ of cases in their population synthesis, compared to the ${\sim}3.1\%$ of cases that formed a hot Jupiter. Moreover, \cite{pet15b} find CHEM produces hot Jupiters at a rate ${\sim}5\%$ per outer cold Jupiter. Given that the fraction of disruptions per hot Jupiter formed is ${\sim}(0.1/3.1){\sim}0.03$, the number of disruptions per star is: ${\sim}0.03{\times}0.05$ hot Jupiters/cold Jupiter ${\times}0.15$ cold Jupiters/star \citep[e.g.][]{cumbutmar08,fulroshir21} ${\sim}2{\times}10^{-4}$ disruptions/star, i.e. 0.023$\%$ of stars should host a desert dweller if they all form via CHEM. This rarity of destruction contrasts with the measured planet occurrence in the desert of ${\sim}10\%$ that of hot Jupiters \citep[occurrence ${\sim}0.1\%$/star;][]{cuiarmhad26}. Nevertheless, the possibility of coplanar HEM can be ascertained for a given desert dweller system by the presence of eccentric ($e{\gtrsim}0.2$), massive (${\gtrsim}1 \ M_{\rm J}$) and distant (${\gtrsim}5 \ \rm au$) planet companions, as predicted by \cite{pet15b}.

While disk migration may deposit planets in the desert's orbital period range \citep[e.g.][assuming the star has shrunk to ${\lesssim}7R_{\odot}$, freeing up orbital periods ${\lesssim}2$ days]{leechi17, hallee20}, it is unclear that disk migration is consistent with desert dwellers' puzzling features, namely, that desert dwellers orbit stars with the same metal-rich distribution as hot Jupiters \citep[][]{visbeh25} and boast exceptionally high densities compared to planets beyond the desert \citep{doyarmacu25}. As photoevaporation alone cannot strip the atmosphere of a mid-sized desert dweller \citep[see e.g. Figure 4 of][and \citealt{ionpav18}]{owelai18}, disk migration is unlikely to allow for enough mass loss to reproduce desert dwellers \citep[unless their low gas fractions arise due to late-time core formation; e.g.][]{lee19}. We therefore consider contamination of aligned desert dwellers via disk migration a remote possibility.

Another alternative for aligned desert dwellers is that they arise via mergers of coplanar smaller planets \citep[e.g.][]{liuhorlin15,chaforohn25}. Since hot small planet occurrence correlates weakly with stellar metallicity \citep[][]{petmarwin18}, such a ``bottom-up" picture may run counter to desert dwellers' preference for metal-rich stars \citep{visbeh25}. Whether small planet \textit{multiplicity} correlates with stellar metallicity remains poorly understood; if so, collisions could indeed be consistent with desert dwellers' preference for metal-rich stars. Future work is needed to estimate how small planet multiplicity scales with stellar metallicity.



\subsection{Frequency of Circumstellar Tori}\label{subsec:tori}

Having clarified the population-level predictions that observations can test to distinguish lossy RLO from HEM, here we consider the prospects for direct observation of lossy RLO in action. As noted by \cite{halmil26}, perhaps the best chance of directly observing hot Jupiters undergoing lossy RLO is a survey akin to the Dispersed Matter Planet Project (DMPP) \citep{hasstabar20,stabarhas26}. In the DMPP, radial velocity follow-up of stars with anomalously low activity indicators -- likely betraying the presence of attenuating circumstellar gas -- is used to discover planets actively losing mass. Our goal in this section is to estimate the number of hot Jupiters undergoing RLO that a survey akin to DMPP is likely to detect \citep[see also][who propose a similar survey]{schmur26}. 

The number of hot Jupiters per star that reach the Roche limit per unit time can be written,

\begin{equation}
    \frac{dN}{dt}\bigg|_{\rm R}=\frac{dN}{d\log P}\bigg|_{\rm R}\frac{d\log P}{dt}\bigg|_{\rm R},
\end{equation}

\noindent where $dN/d\log P\big|_{\rm R}$ denotes hot Jupiter occurrence per logarithmic interval in orbital period, $\big |_{\rm R}$ indicates evaluation at the Roche limit, and $d\log P/dt\big|_{\rm R}$ is the change in logarithmic period experienced by the population per unit time. Under (stellar) tidal transport, $d\log P/dt{=}d\log a/d t{\times}d\log P/d\log a{=}1.5/t_{\rm decay}$ with $t_{\rm decay}{=}a/|\dot{a}_{\rm tid}|{\sim}t_{\rm tid}$ the tidal decay timescale (Equation \ref{equation:ttid}). Over a time interval $\Delta \tau$, the total number of hot Jupiters reaching the Roche limit is,

\begin{align}\label{equation:NHJ}
\begin{split}
    N&\sim \frac{3}{2}\frac{dN}{d\log P}\bigg|_{\rm R}\frac{\Delta \tau}{t_{\rm decay}}\\
    &\sim 10^{-7}\bigg(\frac{\Delta \tau/t_{\rm decay}}{10^{-4}}\bigg) \ \rm planets/star,
\end{split}
\end{align}

\noindent where we have used an occurrence $dN/d\log P\big|_{\rm R}{=}0.03$ planets per 100 stars per 0.25 dex $P$ interval \citep[measured at $P{\sim}1$ day;][]{petmarwin18}. The ratio of timescales accounts for the ${\sim}10^{5}$ yr hot Jupiters spend undergoing lossy RLO; a present-day survey may find hot Jupiters that initiated RLO in the past $10^{5}$ yr such that $\Delta \tau/t_{\rm decay}{\sim}10^{5}/10^{9}{\sim}10^{-4}$. 

Equation \ref{equation:NHJ} indicates that catching a hot Jupiter in the midst of catastrophic RLO is rare, and would require a search of ${\sim}10^{7}$ stars.\footnote{Over ${\sim}$Gyr, Equation \ref{equation:NHJ} indicates ${\sim}10^{-3}$ hot Jupiters/star have cumulatively reached the Roche limit, which encouragingly is comparable to the observed desert dweller occurrence rate \citep[][]{cuiarmhad26}.} Adapting an FGK stellar density $n_{\star}{\sim}10^{-2} \ \rm pc^{-3}$ \citep[see Table 1 in][]{bov17}, we therefore expect ${\sim}$1 hot Jupiter currently undergoing RLO in the nearest $d{\sim}[10^{7}/(4\pi n_{\star}/3)]^{1/3}{\sim}600 \ \rm pc$. Such distances \textit{do} lie within reach of radial velocity instruments \citep[assuming stellar granulation noise does not mask the signal at ${\lesssim}1$ day;][]{stabarhas26}, should stellar activity surveys identify a system undergoing RLO \citep[see e.g. Table 2.1 of][for a compilation of catalogs]{sta18}. Overall, DMPP-style surveys will likely be dominated by hot sub-Neptunes given their ${\sim}10{\times}$ higher occurrence rate than ultrahot Jupiters undergoing orbital decay \citep[e.g.][]{petmarwin18,zinharchr23}.

\section{Conclusion}\label{sec:conclusion}

Recent observational \citep[][]{visbeh25} and theoretical \citep[][]{halmil26} work suggests that the sub-Jovian desert is populated with the remnants of destroyed hot Jupiters. Here, we build on the \cite{halmil26} picture by which hot Jupiters' stripped remains backfill the desert via ``lossy" Roche lobe overflow. Motivated by the fact that 
``lossy" RLO naturally arises due to magnetospheric accretion of angular momentum by the host star, our goal is to determine how mass/angular momentum transfer manifests in the stellar obliquity distribution. Our results guide observations that may test and differentiate the \cite{halmil26} theory from others (chiefly, tidal disruption during HEM).  
Our main conclusions are as follows:

\begin{itemize}
\item Lossy RLO tilts host stars into spin/orbit alignment. Regardless of initial conditions and the details of angular momentum transport, post-mass transfer obliquities cluster within ${\lesssim}60^{\circ}$.
\item Spin/orbit alignment differentiates lossy RLO from tidal disruption during HEM, since post-HEM realignment is inefficient. Contrary to HEM, RLO also predicts an empty sub-Jovian desert around stars above the Kraft break, that desert dwellers are ${\sim}$Gyrs old, and that they may harbor nearby companions.
\item Spin/orbit alignment produced by RLO can be reversed $\textit{after}$ RLO by companion planets. Such companions must lie within ${\lesssim}$2 au, and should be misaligned with the desert dweller.
\item Stars can emerge from RLO slowly rotating, but only if the system begins strongly retrograde. Slowly rotating host stars relieve tension between top-down formation and the prototypical desert dweller, LTT 9779 b.
\item Surveys of nearby stars for ablating planets are more likely to find evaporating hot Neptunes than hot Jupiters actively undergoing catastrophic RLO; nevertheless, such hot Jupiters should exist within ${\sim}$600 pc, amenable to radial velocity discovery.
\end{itemize}

Observations confirming our predictions would not only reveal the fate of hot Jupiters, but open a window into their exposed interiors.

\acknowledgments

We thank the referee for a constructive and helpful report that improved our paper. T.H. thanks the organizers of the Exoplanets and Planet Formation conference at Tsung Dao Lee Institute, Shanghai, China where this work was largely conceived. T.H. also thanks James Jenkins for stimulating discussion on obliquity measurements. The authors acknowledge the MIT Office of Research Computing and Data for providing high performance computing resources that have contributed to the research results reported in this paper. 

\software{Numerical integrations were carried out with \texttt{scipy} \citep[][]{virgomoli20} and \texttt{numpy} \citep[][]{har20}, while plots were created using \texttt{matplotlib} \citep[][]{hun07}.}

\newpage

\appendix{}\label{sec:appendix}

\section{Is ``lossy" Roche lobe overflow adiabatic?}\label{sec:adiabat}

The obliquity predictions made in this work rely on \cite{halmil26}'s ``lossy" RLO mechanism to destroy hot Jupiters and populate the sub-Jovian desert. To calculate the outcome of RLO, \cite{halmil26} employed a streamlined methodology to couple planets' structural/thermal/mass loss/orbital evolution \citep[as outlined in][]{halmil26b}. These coupled structure/mass loss/orbital calculations are simplified in the sense that the planet is assumed to remain in hydrostatic and thermal equilibrium at all times; this equilibrium assumption enables one to follow planetary evolution during RLO using a single ordinary differential equation in planet entropy (see Equation \ref{equation:dsdt} below) rather than the coupled partial differential equations of planetary structure/evolution (i.e. the stellar structure/evolution equations). The purpose of this Appendix is to confirm that this simplified approach to following RLO -- and therefore predicting stellar obliquities --  is accurate, by benchmarking the result against full-fledged hydrodynamics simulations of the planet structure.

We begin by recapping the method used in this paper and \cite{halmil26} to compute RLO evolution. Following standard methodology for planets passively cooling \citep[e.g.][]{hub77,arrbil06}, \cite{halmil26} evolved the entropy in their planets' innermost convective zones under radiative cooling alone (``following the adiabats"). Using a pre-computed grid of planet structures in hydrostatic and thermal equilibrium, \cite{halmil26} evolved the following equation for innermost convective zone entropy $S$ \citep[e.g.][]{marcum14}: 

\begin{equation}\label{equation:dsdt}
    \frac{dS}{dt}=\frac{-L}{\int_{\rm conv}Tdm},
\end{equation}

\noindent where $L$ is the planet's cooling luminosity (assumed spatially constant) and $\int_{\rm conv}Tdm$ the integral of temperature over the convective zone mass. Equation \ref{equation:dsdt} describes the decrease in entropy solely due to radiative cooling. \cite{halmil26} justified their use of \ref{equation:dsdt} on the basis that their planets' convective zones remained isentropic \citep[convective time ${\ll}$ mass loss time; e.g.][]{pacsie72} and $L$ remained spatially constant \citep[thermal relaxation time in the radiative zone ${<}$ mass loss time; e.g.][]{arrbil06}. \cite{halmil26} found that lossy RLO occurred so quickly that it was ``adiabatic", i.e. $m_{\rm p}/|\dot{m}_{\rm p}|{\ll}S/|\dot{S}|$ \citep[as found in other planetary mass loss contexts; e.g.][]{barselcha04}. We seek to determine whether the planet's entropy (and therefore structure) changes during mass transfer in a more realistic calculation.

To compare the ``following the adiabats" method of computing planet structure/thermal/mass loss evolution against hydrodynamic simulations, we use \texttt{MESA} version 25.12.1 \citep[][]{paxbildot11,paxcanarr13,paxmarsch15,paxschbau18,paxsmosch19} with hydrodynamics enabled. Using \texttt{MESA}, we are able to track energy changes at each atmospheric layer as gas parcels advect, experience $PdV$ work, undergo irradiation heating, etc. \citep[see e.g.][]{owewu16}. Our $\texttt{MESA}$ files are provided in a Zenodo catalog.\footnote{\href{https://doi.org/10.5281/zenodo.19008881}{Available here}.}

\begin{figure*}
\centering
\includegraphics[width=0.75\textwidth]{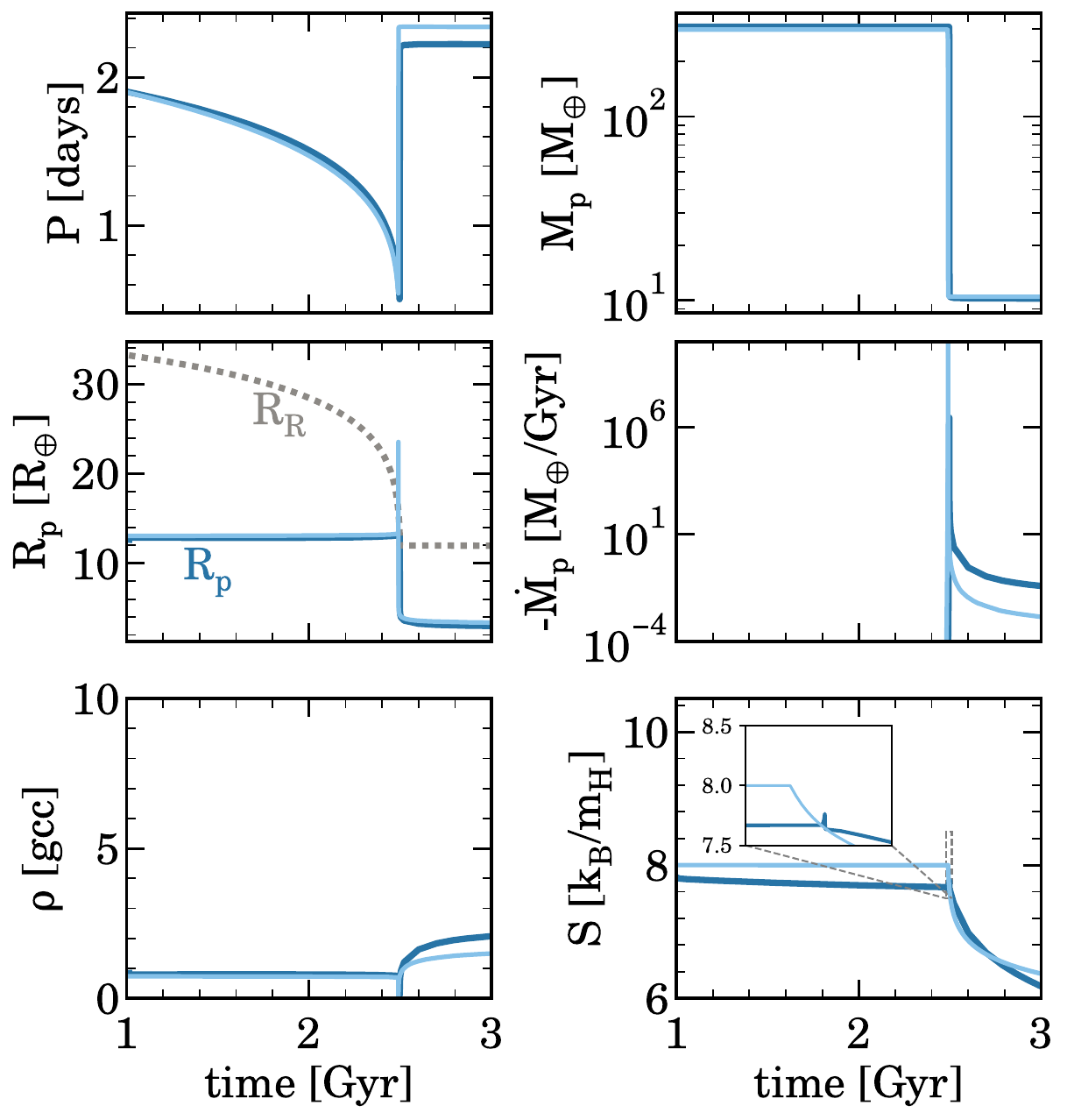}
\caption{``Lossy" RLO of a hot Jupiter with a 10 $M_{\oplus}$ core as computed two different ways. Top to bottom, left panels: orbital period, photospheric radius $\rm R_{\rm p}$ ($\rm R_{\rm R}$ is the Roche radius), and bulk density. Top to bottom, right panels: mass, mass loss rate, and entropy in the convective zone (innermost cell in $\texttt{MESA}$ model). Light blue curves use the \cite{halmil26,halmil26b} ``following the adiabats" structure/thermal evolution model, while dark blue use \texttt{MESA} with hydrodynamics enabled. Regardless of planetary structure/thermal evolution details, lossy RLO robustly destroys hot Jupiters (and produces desert dwellers). 
\label{figure:mesacomp}}
\end{figure*}

\begin{figure*}
\centering
\includegraphics[width=0.75\textwidth]{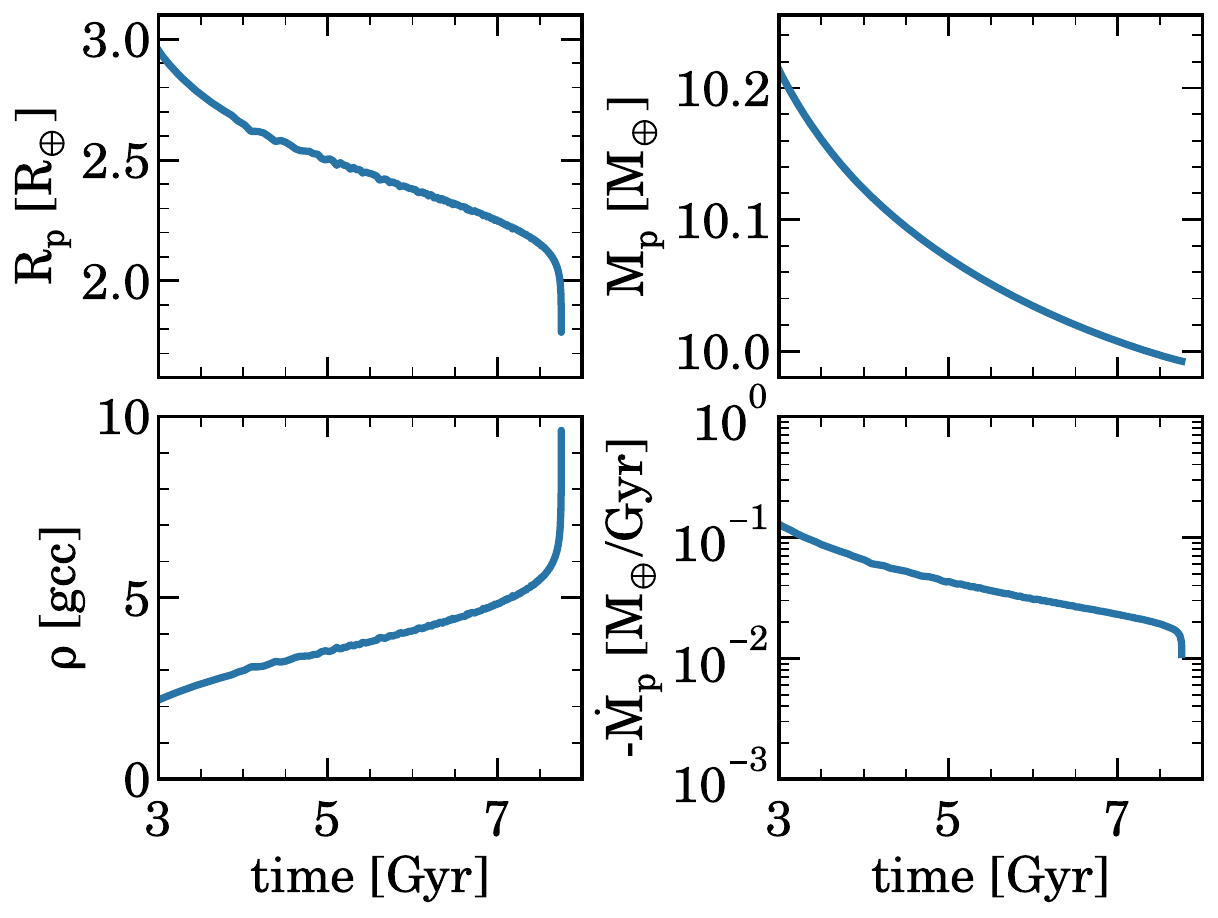}
\caption{Post-RLO evolution of the planet from Figure \ref{figure:mesacomp} (note the differing $xy$ axes). Our \texttt{MESA} calculation produces late-time mass loss which evacuates the remaining ${\sim}1{\%}$ of hydrogen left after peak RLO, predicting denser desert dwellers than \cite{halmil26}. This late-time behavior reflects the transition from Roche lobe overflow to photoevaporation, aided by the planet's structure becoming radiative. 
\label{figure:mesacomp2}}
\end{figure*}

Figure \ref{figure:mesacomp} compares RLO evolution using the two different approaches. The evolution is generally in strong agreement between methods. As displayed in Figure \ref{figure:mesacomp2}, $\texttt{MESA}$ produces complete atmospheric evacuation in the ${\sim}$Gyr following RLO, whereas the \cite{halmil26} calculations maintain a thin ${\sim}1\%$ by mass atmosphere (see e.g. their Figure 5). This late-stage atmospheric mass loss occurs as RLO gives way to photoevaporation \cite[implemented in $\texttt{MESA}$ following][]{halmil26}. We have traced this difference in final mass to the transition from a convective to a radiative structure as the atmosphere becomes extremely thin (${\lesssim}10^{-1} \ M_{\oplus}$). As a radiative window begins to open at depth, the internal temperature gradient flattens (or becomes weakly inverted) while the entropy in the interior rises and luminosities become negative; the envelope expands under $PdV$ work, producing an inwardly-directed radiative flux. The planet remains susceptible to photoevaporation until the remnant atmosphere is fully evacuated. 

Another difference between calculation approaches is reflected in entropy evolution during mass loss (see zoomed inset in Figure \ref{figure:mesacomp}). The $\texttt{MESA}$ run shows that the entropy in the convective zone (measured at the model's bottom cell, corresponding to the base of the innermost convective zone) is $\textit{not}$ fixed during mass transfer, but rather mirrors the mass loss rate (growing to a peak before declining). The evolution in entropy can be qualitatively understood from the ratio of escaping gas' specific gravitational potential (${-}GM_{\rm p}/R_{\rm p}$) to internal (denoted $u{\sim}c^{2}_{\rm s}$, with $c_{\rm s}$ the sound speed) energy. If material is only marginally bound ($|GM_{\rm p}/R_{\rm p}|{\gtrsim}u$; i.e. $R_{\rm p}{\sim}R_{\rm Bondi}{=}GM_{\rm p}/2c^{2}_{\rm s}$) mass loss does not remove very much negative energy from the planet; losing mass while maintaining a very negative total energy requires entropy decrease. Likewise, losing strongly bound material ($|GM_{\rm p}/R_{\rm p}|{\gg}u$) removes significant amounts of negative energy so that the total planet energy will become very much less negative; in the case of RLO, mass loss is so intense that entropy grows. The planets in Figure \ref{figure:mesacomp} satisfy $|GM_{\rm p}/R_{\rm p}|{\gg}u$ initially ($R_{\rm p}{\ll}R_{\rm Bondi}$), so that entropy grows upon RLO. After peak mass loss, entropy declines as the planet transitions into the $|GM_{\rm p}/R_{\rm p}|{\gtrsim}u$ regime. Entropy then grows again once the atmosphere becomes radiative.

We note that the two approaches also differ in computation time: our $\texttt{MESA}$ run took ${\sim}$4 days using 40 cores ($\texttt{OMP$\_$NUM$\_$THREADS}{=}$40), whereas the \cite{halmil26} method takes ${\sim}$30 minutes on a single core. 

\bibliography{hallatt}{}
\bibliographystyle{aasjournal}

\end{document}